\include{graphicsx}

\documentclass[preprint]{aastex}
\usepackage{emulateapj5}

\slugcomment{Submitted 07/22/02, accepted 12/18/02, to appear in the April 10, 2003 issue of the Astrophysical Journal}

\shorttitle{Very Low Mass Binaries}
\shortauthors{Close et al.}

\begin{document}


\title{Detection of Nine M8.0-L0.5 Binaries: The Very Low Mass Binary Population and its Implications for Brown Dwarf and VLM Star Formation}


\author{Laird M. Close$^1$, Nick Siegler$^1$, Melanie Freed$^1$, \& Beth Biller$^1$}

\email{lclose@as.arizona.edu}

\affil{$^1$Steward Observatory, University of Arizona, Tucson, AZ 85721}
%\affil{$^2$Institute for Astronomy, University of Hawaii, Honolulu, HI}
%\affil{$^3$European Southern Observatory, Garching, Germany}   



\begin{abstract} 

Use of the highly sensitive Hokupa'a/Gemini curvature wavefront sensor
has allowed direct adaptive optics (AO) guiding on
very low mass (VLM) stars with SpT=M8.0-L0.5. A survey of 39 such
objects detected 9 VLM binaries (7 of which were discovered for the
first time to be binaries). Most of these systems are tight
(separation $<5$ AU) and have similar masses ($\Delta Ks<0.8$ mag;
$0.85<q<1.0$). However, 2 systems (LHS 2397a and 2M2331016-040618) have large
$\Delta Ks>2.4$ mag and consist of a VLM star orbited by a much
cooler L7-L8 brown dwarf companion. Based on this flux limited
($Ks<12$ mag) survey of 39 M8.0-L0.5 stars (mainly from the 2MASS sample of
\cite{giz00}) we find a sensitivity corrected binary fraction in the range $15\pm7\%$ for
M8.0-L0.5 stars with separations $>2.6$ AU. This is slightly less than
the $32\pm9\%$ measured for more massive M0-M4 dwarfs over the same
separation range \citep{fis92}. It appears M8.0-L0.5 binaries (as well
as L and T dwarf binaries) have a much smaller semi-major axis
distribution peak ($\sim 4$ AU) compared to more massive M and G
dwarfs which have a broad peak at larger $\sim 30$ AU separations. We
also find no VLM binary systems (defined here as systems with
$M_{tot}<0.185 M_{\sun}$) with separations $>15$ AU.

We briefly explore possible reasons why VLM binaries are slightly less
common, nearly equal mass, and much more tightly bound compared to
more massive binaries.  We investigate the hypothesis that the lack of
wide ($a>20$ AU) VLM/brown dwarf binaries may be explained if the
binary components were given a significant differential velocity kick. Such a
velocity kick is predicted by current ``ejection'' theories, where
brown dwarfs are formed because they are ejected from their embryonic
mini-cluster and therefore starved of accretion material. We find that
a kick from a close triple or quadruple encounter (imparting a
differential kick of $\sim 3$ km/s between the members of an escaping
binary) could reproduce the observed cut-off in the semi-major axis
distribution at $\sim20$ AU. However, the estimated binarity ($\la
5\%$; \cite{bat02}) produced by such ejection scenarios is below the
$15\pm7\%$ observed. Similarly, VLM binaries could be the final hardened
binaries produced when a mini-cluster decays. However, the models of
\cite{ste98, dur01} also cannot produce a VLM binary fraction above $\sim
5\%$. The observed VLM binary frequency could possibly be produced by
cloud core fragmentation. Although, our estimate of a
fragmentation-produced VLM binary semi-major axis distribution
contains a significant fraction of
``wide'' VLM binaries with $a>20$ AU in contrast to observation. In
summary, more detailed theoretical work will be needed to explain
these interesting results which show VLM binaries to be a
significantly different population from more massive M
\& G dwarf binaries.

\end{abstract}


\section {Introduction}

Since the discovery of Gl 229B by \cite{nak95} there has been intense
interest in the direct detection of brown dwarfs and very low mass
(VLM) stars and their companions. According to the current models of
\cite{bur00} and \cite{cha00}, stars with spectral types of M8.0-L0.5
will be just above the stellar/substellar boundary. However, modestly
fainter companions to such primaries could themselves be
substellar. Therefore, a survey of M8.0-L0.5 stars should detect
binary systems consisting of VLM primaries with VLM or brown dwarf
secondaries.

The binary frequency of M8.0-L0.5 stars is interesting in its own right since little is known about how common M8.0-L0.5 binary systems are. It is
not clear currently if the M8.0-L0.5 binary separation distribution is
similar to that of M0-M4 stars; in fact, there is emerging evidence
that very low mass L \& T dwarf binaries tend to have smaller separations and possibly lower binary frequencies compared to more massive M and G stars \citep{mar99, rei01a, bur03}. 

%Despite the strong interest in such very low mass binaries, only 9 such
%systems have been detected to date. In our previous paper, \cite{clo02a}, a de%tailed history of each these detections is given in the introduction. 

Despite the strong interest in such very low mass (VLM) binaries
($M_{tot} < 0.185 M_{\sun}$), only 24 such systems are known (see
Table \ref{tbl-3} for a complete list). A brief overview of these
systems starts with the first double L dwarf system which was imaged
by HST/NICMOS by \cite{mar99}. A young spectroscopic binary brown
dwarf (PPL 15) was detected in the Pleiades (\cite{bas99}) but this
spectroscopic system is too tight to get separate luminosities for
each component. A large HST/NICMOS imaging survey by \cite{mar00} of
VLM dwarfs in the Pleiades failed to detect any brown dwarf binaries
with separations $>0.2\arcsec$ ($\ga 27$ AU). Detections of nearby field
binary systems were more successful. The nearby object Gl 569B was
resolved into a $0.1\arcsec$ (1 AU) binary brown dwarf at Keck and the
6.5m MMT (\cite{mar99b};
\cite{ken01}; \cite{lan01}). Keck seeing-limited NIR 
imaging marginally resolved two more binary L stars (\cite{kor99}). A
survey with WFPC2 detected four more (three newly discovered and one confirmed from \cite{kor99}) tight equal
magnitude binaries out of a sample of 20 L dwarfs
(\cite{rei01a}). From the same survey \cite{rei02} found a M8 binary (2M1047; later discovered independently by our survey). Guiding on HD130948 with adaptive optics (AO),
\cite{pot02a} discovered a companion binary brown dwarf system. Recently,
\cite{bur03} have detected two T dwarf binaries with HST. Finally, 12
more L dwarf binaries have been found by analyzing all the currently
remaining HST/WFPC2 data collected on L dwarfs \citep{bou02}. Hence,
the total number of binary VLM stars and brown dwarfs currently known
is just 24. Of these, all but one have luminosities known for
each component, since one is a spectroscopic binary.

Here we have carried out our own high-spatial resolution binary survey of VLM stars employing adaptive optics. In total, we have detected 12 VLM binaries (10 of these are new
discoveries) in our AO survey of 69 VLM stars. Three of these systems
(LP415-20, LP475-855, \& 2MASSW J1750129+442404) have M7.0-M7.5
spectral types and are discussed in detail elsewhere
\citep{sie02}. In this paper, we discuss the remaining 9 cooler
binaries with M8.0-L0.5 primaries detected in our survey (referred
herein as M8.0-L0.5 binaries even though they may contain L1-L7.5
companions; see Table \ref{tbl-1} for a complete list of these
systems). 

Two of these systems (2MASSW J0746425+200032 and 2MASSW
J1047127+402644) were in our sample but were previously imaged in
the visible by HST and found to be binaries (\cite{rei01a,rei02}). Here we present the first resolved IR observations of these two
systems and new astrometry. The seven remaining systems were
all discovered to be binaries during this survey. The first four systems
discovered in our survey (2MASSW J1426316+155701, 2MASSW
J2140293+162518, 2MASSW J2206228-204705, and 2MASSW J2331016-040618)
have brief descriptions in \cite{clo02b}.  However, we have
re-analyzed the data from
\cite{clo02b} and include it here for completeness with slightly
revised mass estimates. The very interesting M8/L7.5 system LHS 2397a
discovered during this survey is discussed in detail elsewhere
\citep{fre02} yet is included here for completeness. The newly
discovered binaries 2MASSW J1127534+741107 and 2MASSW J1311391+803222
are presented here for the first time.

These nine M8.0-L0.5 binaries are a significant addition to the other very low mass M8-T6 binaries known to date listed in Table \ref{tbl-3} 
\citep{bas99,mar99,kor99,rei01a,lan01,pot02a,bur03,bou02}. With relatively short periods our
new systems will likely play a significant role in the
mass-age-luminosity calibration for VLM stars and brown dwarfs. It is
also noteworthy that we can start to characterize this new population
of M8.0-L0.5 binaries. We will outline how VLM binaries are
different from their more massive M and G counterparts. Since VLM binaries are so tightly bound we hypothesize that very little dynamical evolution of the distribution has occurred since their formation. We attempt to
constrain the formation mechanism of VLM stars and brown dwarfs from
our observed semi-major axis distribution and binarity statistics.

%%Moreover, we show for the first time
%%that it possible to guide {\it directly} on a low mass star/brown
%%dwarf with adaptive optics on an 8m telescope, opening up a new
%%technique for diffraction-limited brown dwarf observations.

%%We will also present
%%vidence of how the semi-major axis distribution of brown dwarf
%%binaries is significantly different to that of stars.

\section{An AO survey of nearby M8.0-L0.5 field stars}

 As outlined in detail in \cite{clo02a}, we utilized the University of
 Hawaii curvature adaptive optics system Hokupa'a
 \citep{gra98,clo98a}, which was a visitor AO instrument on the Gemini
 North Telescope. This highly sensitive curvature AO system is well
 suited to locking onto nearby, faint ($V\sim20$), red ($V-I>3$), M8.0-L0.5 stars to
 produce $\sim 0.1\arcsec$ images (which are close to the $0.07\arcsec$
 diffraction-limit in the $K^\prime$ band). We can guide on such faint ($I\sim 17$) targets with a curvature AO system (such as Hokupa'a) by utilizing its zero-read noise wavefront sensor (for a detailed explanation of how this is possible see \cite{sie02a}). We utilized this unique
 capability to survey the nearest extreme M and L stars (M8.0-L0.5) to
 characterize the nearby VLM binary population.

Here we report the results of all our Gemini observing runs in 2001
and 2002. We have observed all 32 
M8.0-M9.5 stars with $Ks<12$ mag from the list of \cite{giz00}. It should be noted that the
M8.0-M9.5 list of \cite{giz00} has some selection constraints:
galactic latitudes are all $> 20$ degrees; and from $0<RA<4.5$ hours
$DEC<30$ degrees; and there are gaps in the coverage due to the past
availability of the 2MASS scans. A bright L0.5 dwarf with
$Ks<12$ was also observed (selected from \cite{kir00}). Six additional
bright ($Ks<12$) M8.0-M9.5 stars were selected from \cite{rei02} and \cite{cru02}. In total 39 M8.0-L0.5
stars have now been imaged at high resolution ($\sim 0.1\arcsec $) with
AO compensation in our survey. For a complete list of these M8-L0.5 target stars see Table \ref{tbl-0} (single stars) and Table \ref{tbl-1} (stars found to be binaries).

%Typically we achieved AO corrected $0.13 \arcsec$ fwhm images in the
%$K^\prime$ ($2.1\micron$) band. Since the survey is on-going we will
%report in more detail about all the objects observed in a more
%comprehensive paper in the near future when all M8.0-L0.5 stars in
%\cite{giz00} have been observed. In this letter we limit the
%discussion to the 4 new binaries 2M1426, 2M2140, 2M2206 and 2M2331 and make ge%neral conclusions about M8.0-L0.5 binaries. 

Nine of our 39 targets were clearly tight binaries
($sep < 0.5 \arcsec$). We observed each of these objects by dithering
over 4 different positions on the QUIRC $1024 \times 1024$ NIR ($1-2.5 \mu m$) detector \citep{hod96} which has a
$0.0199\arcsec$/pixel plate scale at Gemini North. At each position we took 3x10s exposures at
J, H, $K^{\prime}$, and 3x60s exposures at H, resulting in unsaturated
120s exposures at J, H, and $K^{\prime}$ with a deep 720s exposure
at H band for each binary system.

\section{Reductions}

We have developed an AO data reduction pipeline in the IRAF language
which maximizes sensitivity and image resolution. This pipeline is
standard IR AO data reduction and is described in detail in
\cite{clo02a}.  

Unlike conventional NIR observing at an alt-az telescope (like Gemini), we
disable the Cassegrain rotator so that the pupil image is fixed w.r.t.
the detector. Hence the optics are not rotating w.r.t. the camera or
detector. In this manner the residual static PSF aberrations are fixed
in the AO images enabling quick identification of real companions as
compared to PSF artifacts. The pipeline cross-correlates and aligns each image,
then rotates each image so north is up and east is to the left, then median
combines the data with an average sigma clip rejection at the $\pm2.5
\sigma$ level. By use of a cubic-spline interpolator the script
preserves image resolution to the $<0.02$ pixel level. Next the custom IRAF script
produces two final output images, one that combines all the images
taken and another where only the sharpest 50\% of the images are
combined. 

\begin{figure}
 \includegraphics[angle=0,width=\columnwidth]{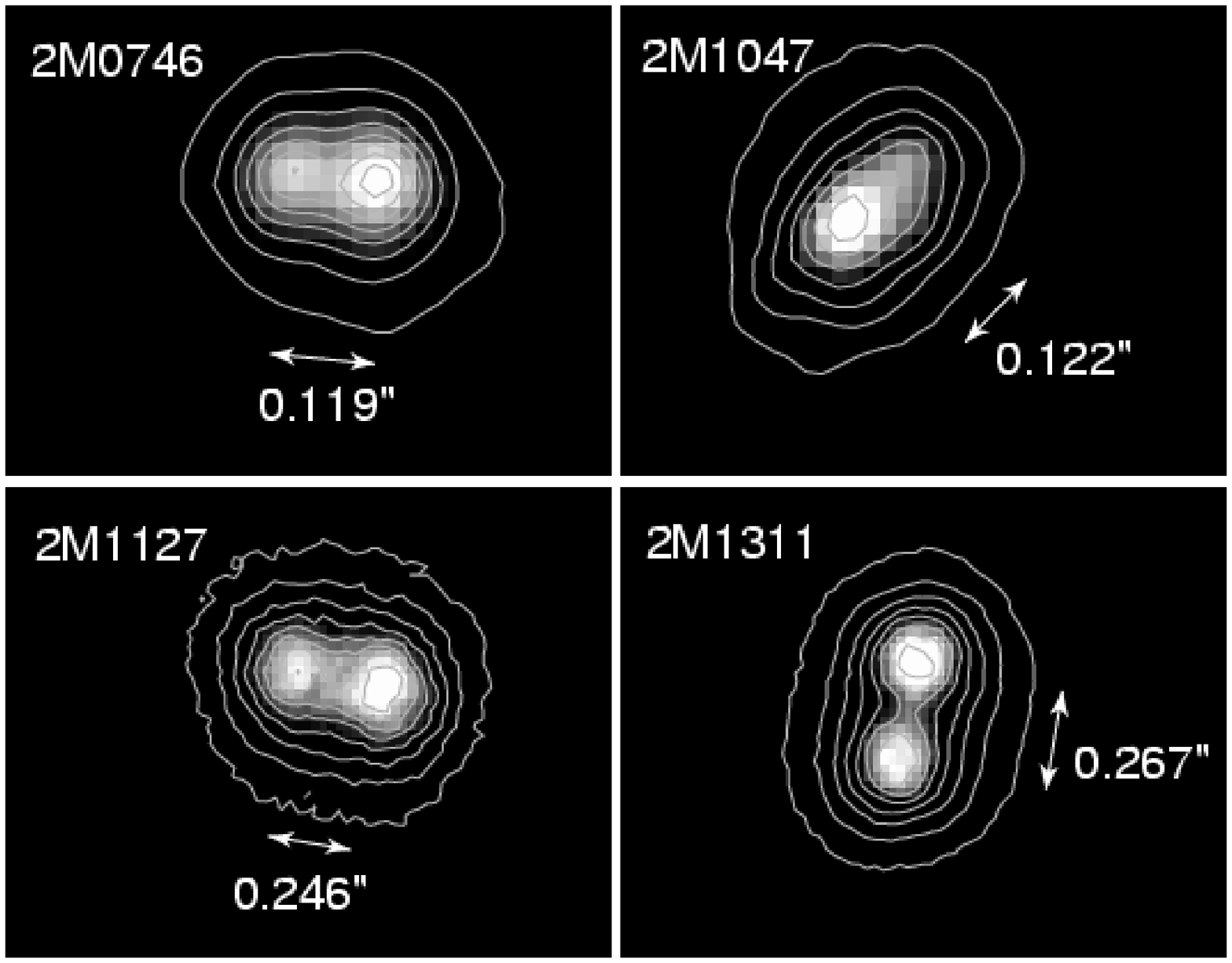}\caption{The 12x10s $K^\prime$ image of the L0.5 2MASSW J0746425+200032
binary (which is the one overlap object between our study and that of
\cite{rei01a}). Note the very different position ($PA=85.7^{\circ}$, $sep=0.121\arcsec= 1.49$ AU on 2002/2/7) of our image
compared to the PA=$168.8^{\circ}$ sep 2.7 AU published by
\cite{bou02} from HST observations of \cite{rei01a} at 2000/4/15. This suggests that the orbit could have a period of $\sim 4 $yr. In the near future this system could have an orbital solution. At a resolution of $<0.1\arcsec$ both components are
clearly visible. We also show $K^\prime$ images of the new binaries
2MASSW J1047127+402644, 2MASSW J1127534+741107, \& 2MASSW
J1311391+803222. The pixels are $0.199\arcsec$pix$^{-1}$. The contours are
linear at the 90, 75, 60, 45, 30, 15, and 1\% levels. North is up
and east left in each image.\label{fig1}}
\end{figure}

\begin{figure}
\includegraphics[angle=0,width=\columnwidth]{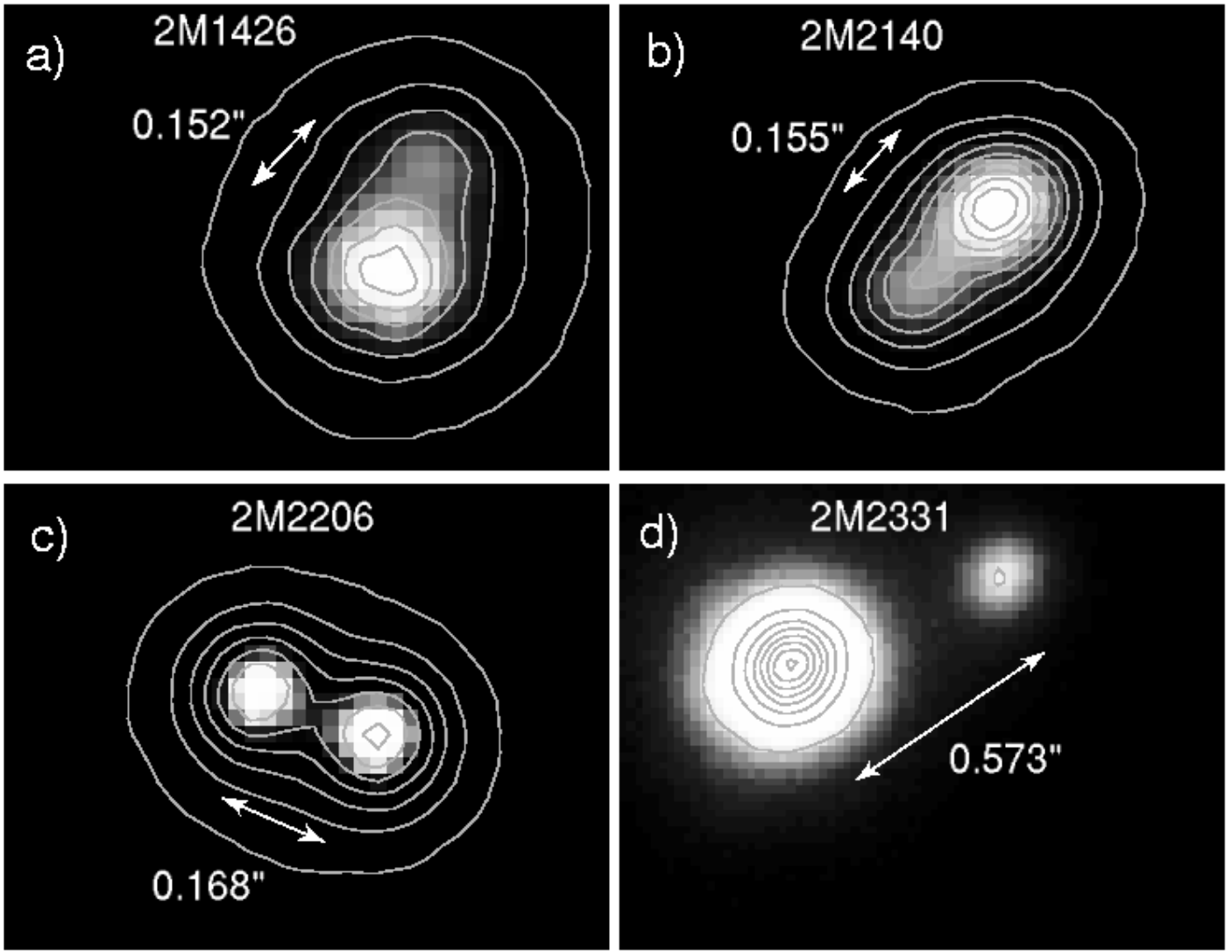}\caption{In figure (a) we see the 12x10s $K^\prime$ image of the 2MASSJ 1426316+155701 binary discussed in \cite{clo02b} at a resolution of
$0.131\arcsec$. In Figures (b-d) we show $K^\prime$ images of the binaries 2MASSW J2140293+162518, 2MASSWJ 2206228-204705, and 2MASSWJ 2331016-040618, respectively. The pixels are
0.0199$\arcsec$pix$^{-1}$. The contours are linear at the 90, 75, 60, 45, 30, 15, and
1\% levels. North is up and east left in each image.\label{fig1b}} \end{figure}

This pipeline produced final unsaturated 120s exposures at J ($FWHM
\sim 0.15\arcsec$) , H ($FWHM \sim 0.14 \arcsec$), and $K^{\prime}$
($FWHM \sim 0.13 \arcsec$) with a deep 720s exposure ($FWHM \sim 0.14
\arcsec$) at H band for each binary system. The dithering produces a
final image of 30x30$\arcsec$ with the most sensitive region
($10\times 10 \arcsec$) centered on the binary. Figures
\ref{fig1} and \ref{fig1b} illustrates $K^{\prime}$ images of each of the systems.

\placefigure{fig1} 
\placefigure{fig2}

In Table \ref{tbl-1} we present the analysis of the images taken of
the 9 new binaries from our Gemini observing runs. The photometry was based on DAOPHOT
PSF fitting photometry \citep{ste87}. The PSFs used were the reduced
12x10s unsaturated data from the next (and previous) single VLM stars
observed after (and before) each binary. The PSF stars always had a
similar IR brightness, a late M spectral type, and were observed at a
similar airmass.  The resulting $\Delta magnitudes$ between the components are listed in
Table \ref{tbl-1}; their errors in $\Delta mag$ are the differences in
the photometry between two similar PSF stars. The individual fluxes were
calculated from the flux ratio measured by DAOPHOT. We made the
assumption $\Delta K^\prime
\sim \Delta Ks$, which is correct to 0.02 mag according the models of
 \cite{cha00}. Assuming $\Delta K^\prime = \Delta Ks$ allows us to use the
2MASS integrated Ks fluxes of the blended binaries to solve for
the individual $Ks$  fluxes of each component (see Table \ref{tbl-2}).

The platescale and orientation of QUIRC was determined from a short
exposure of the Trapezium cluster in Orion and compared to published
positions as in \cite{sim99}. From these observations a platescale of
$ 0.0199\pm 0.0002\arcsec$/pix and an orientation of the Y-axis
($0.3\pm 0.3$ degrees E of north) was determined. Astrometry for
each binary was based on the PSF fitting. The astrometric errors were
based on the range of the 3 values observed at J, H, and $K^\prime$ and
the systematic errors in the calibration added in quadrature.

%%The Astrometry was based on lightly (100 iterations) LUCY deconvolve
%%data from the 12x5s unsaturated H image (see Figure
%%\ref{fig1}). For the
%%deconvolution we used the same PSF as above. A good
%%deconvolution result was obtained with a restored resolution of
%%$FWHM=0.080\arcsec$. This high resolution allowed for excellent
%%astrometric accuracy to be achieved by centroiding on the well separated
%%stars. These astrometric measurements were also
%%checked against the PSF fitting photometry of DAOPHOT and were found
%%to be in excellent agreement (DAOPHOT gave $separation=0.155\arcsec$,
%%$PA=343.5^{o}$).

\placetable{tbl-1}

\placetable{tbl-2}

\section{Analysis}

\subsection {Are the companions physically related to primaries?}

%We believe there is a very high probability that all three companions
%are physically associated with their primaries. This is due to the
%small space density of very red ($J-Ks>1.0$) background objects in the
%field. 

Since \cite{giz00} only selected objects $>20$ degrees above the
galactic plane, we do not expect many background late M or L stars in
our images. In the $3.6$x$10^4$ square arcsecs already surveyed, we have
not detected a very red $J-Ks>0.8$ mag background object in any of the
fields. Therefore, we estimate the probability of a chance projection
of such a red object within $<0.5\arcsec$ of the primary to be
$<2$x$10^{-5}$. Moreover, given the rather low space density ($0.0057 \pm 0.0025
pc^{-3}$) of L dwarfs \citep{giz01}, the probability of a background L dwarf within
$<0.5\arcsec$ of any of our targets is $<10^{-16}$. We
conclude that all these very red, cool objects are physically related
to their primaries.

\subsection {What are the distances to the binaries?}

Unfortunately, there are no published trigonometric parallaxes for six of
the nine systems. The three systems with parallaxes are: 2M0746 \citep{dah02}; LHS 2397a \citep{van95}; and
2M2331 (associated with a Hipparcos star HD221356
\cite{giz00}). For the remaining six systems without parallaxes we can estimate the distance based on the trigonometric parallaxes of other
well studied M8.0-L0.5 stars from \cite{dah02}. The distances of all
the primaries were determined from the absolute Ks magnitudes (using
available 2MASS photometry for each star with trigonometric parallaxes
from \cite{dah02}), which can be estimated by $M_{Ks}=7.71+2.14(J-Ks)$
for M8.0-L0.5 stars
\citep{sie02}. This relationship has a $1 \sigma$ error of 0.33 mag
which has been added in quadrature to the J and Ks photometric errors
to yield the primary component's $M_{Ks}$ values in Table \ref{tbl-2} and plotted as crosses in Figures \ref{fig2} - \ref{fig2i}.  As can be seen from Table \ref{tbl-2} all but one of our systems are
within 29 pc (the exception is 2M1127 at $\sim33$ pc).

\subsection {What are the spectral types of the components?}

We do not have spatially resolved spectra of both components in any of
these systems; consequently we can only try to fit the $M_{Ks}$ values
in Table \ref{tbl-2} to the relation $SpT=3.97M_{Ks}-31.67$ which is
derived from the dataset of \cite{dah02} by
\cite{sie02}. Unfortunately, the exact relationship between $M_{Ks}$
and VLM/brown dwarf spectral types is still under study.  It is
important to note that these spectral types are only a guide since the
conversion from $M_{Ks}$ to spectral type carries at least $\pm1.5$
spectral subclasses of uncertainty. Fortunately, none of the following
analysis is dependent on these spectral type estimates.

It is interesting to note that six of these secondaries are likely L
dwarfs. In particular 2M2331B is likely a L7 and LHS 2397aB is likely a
L7.5. Both 2M2331B and
LHS 2397aB are very cool, late L companions.

\placefigure{fig2}

\subsection {What are ages of the systems?}

Estimating the exact age for any of these systems is difficult since
there are no Li measurements yet published (which could place an upper
limit on the ages). An exception to this is LHS 2397a for which no Li
was detected \cite{mar94}. For a detailed discussion on the age of LHS
2397a see
\cite{fre02}. For each of the remaining systems we have conservatively assumed that the
whole range of common ages in the solar neighborhood (0.6 to 7.5 Gyr) may
apply to each system \citep{cal99}. However, \cite{giz00} observed very low proper motion
($V_{tan}<10$ km/s) for the 2M1127, 2M2140, and 2M2206 systems.
%%($\mu RA=0.108\arcsec/yr$, $\mu DEC=-0.056\arcsec/yr$)
%%gives $V_{tan}=12$ km/s. 
%%Unfortunately the low resolution of the
%%spectrum obtained by \cite{giz00} does not allow the radial velocity
%%(and hence full 3-dimensional space motion) to be calculated at this
%%time. 
These three systems are among the lowest velocity M8's in the entire survey of
\cite{giz00} suggesting a somewhat younger age since these systems have
not yet developed a significant random velocity like the other older ($\sim5$ Gyr) M8.0-L0.5
stars in the survey. 
%%Moreover, the low 1.2 Angstrom EW of $H_{\alpha}$
%%may also suggest a youthful age for such a late spectral type (where
%%$H_{\alpha}$ appears more strongly in older systems;
%%\cite{giz00}). Indeed, 2M1426 occupies a uniquely young position as
%%the lowest $H_{\alpha}$ and lowest $V_{tan}$ of any of the 14 M8.0-L0.5
%%objects in the \cite{giz00} survey. 
Therefore, we assign a slightly younger age of $3.0^{+4.5}_{-2.4}$ Gyr
to these 3 systems, but leave large error bars allowing ages from
0.6-7.5 Gyr ($\sim 3$ Gyr is the maximum age for the kinematically young
stars found by \cite{cal99}).  The other binary systems 2M0746,
2M1047, 2M1311, and 2M2331 appear to have normal $V_{tan}$ and are
more likely to be older systems. Hence we assign an age of
$5.0_{-4.4}^{+2.5}$ Gyr to these older systems \citep{cal99}. It
should be noted that there is little significant difference between the
evolutionary tracks for ages $1-10$ Gyr when $SpT<L0$ 
\citep{cha00}. Therefore, the exact age is not absolutely critical to
estimating the approximate masses for M8.0-L0.5 stars (see Figure
\ref{fig2}).

\subsection {The masses of the components}

\begin{figure}
\includegraphics[angle=90,width=\columnwidth]{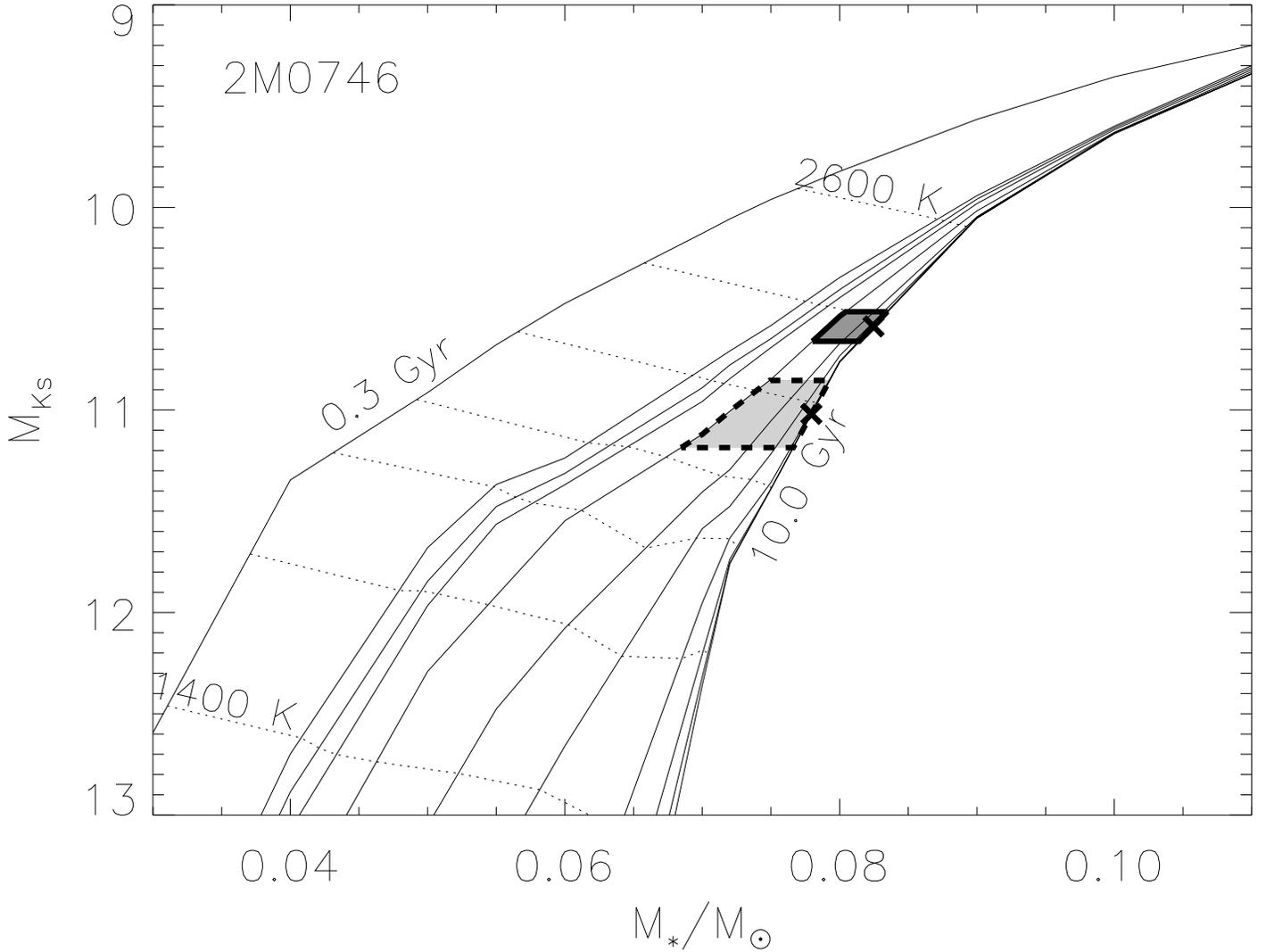}\caption{The
latest \cite{cha00} DUSTY evolutionary models are shown custom integrated over
the Ks bandpass. The locations of the 2 components of 2M0746 are
indicated by the crosses. The polygon encloses the error in the
$M_{Ks}$ and age range $0.6-7.5$ Gyr of the system. The $M_{Ks}$ of
each secondary is determined by the addition of $\Delta Ks$ plus the
$M_{Ks}$ of the primary. The errors for the primary are enclosed by
the upper polygon (outlined by a solid thick line), the secondary is
bounded by the lower polygon (outlined by a thick dashed line). The
models suggest a primary mass of 0.082 $M_{\sun}$ with a range $
0.078-0.083 M_{\sun}$ with temperatures of 2375 K (2323-2409K). For
the secondary the models suggest a mass of $0.078 M_{\sun}$ with a
range $ 0.068-0.079 M_{\sun}$ with temperatures of 2173K (2034-2252
K). The isochrones plotted are 0.3, 0.6, 0.65, 0.7, 0.85, 1.2, 1.7, 3.0, 5.0, 7.5, \& 10.0 Gyr. The isotherms run in equal intervals from 2600 K to 1400 K in steps of 200 K.  \label{fig2}}
\end{figure}

\begin{figure}
\includegraphics[angle=90,width=\columnwidth]{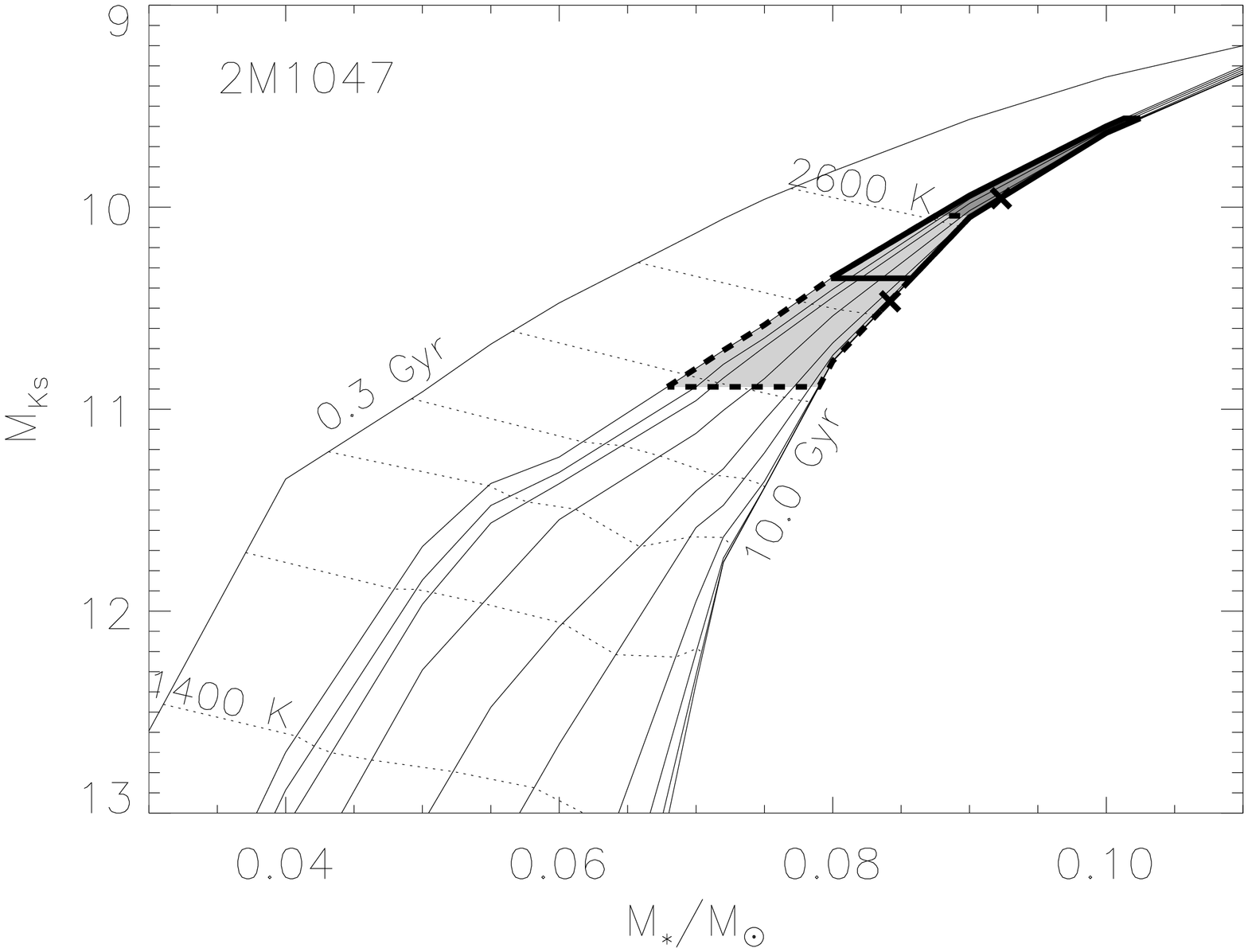}\caption{The same as figure 3 except for 2M1047. Note the larger errors in $M_{Ks}$ compared to 2M0746 since we have to currently rely on a photometric parallax for this system. The
models suggest a primary mass of 0.092 $M_{\sun}$ with a range $
0.079-0.102 M_{\sun}$ with temperatures of 2660 K (2460-2830K). For
the secondary the models suggest a mass of $0.084 M_{\sun}$ with a
range $ 0.068-0.090 M_{\sun}$ with temperatures of 2430K (2164-2625
K). \label{fig2b}}
\end{figure}

\begin{figure}
\includegraphics[angle=0,width=\columnwidth]{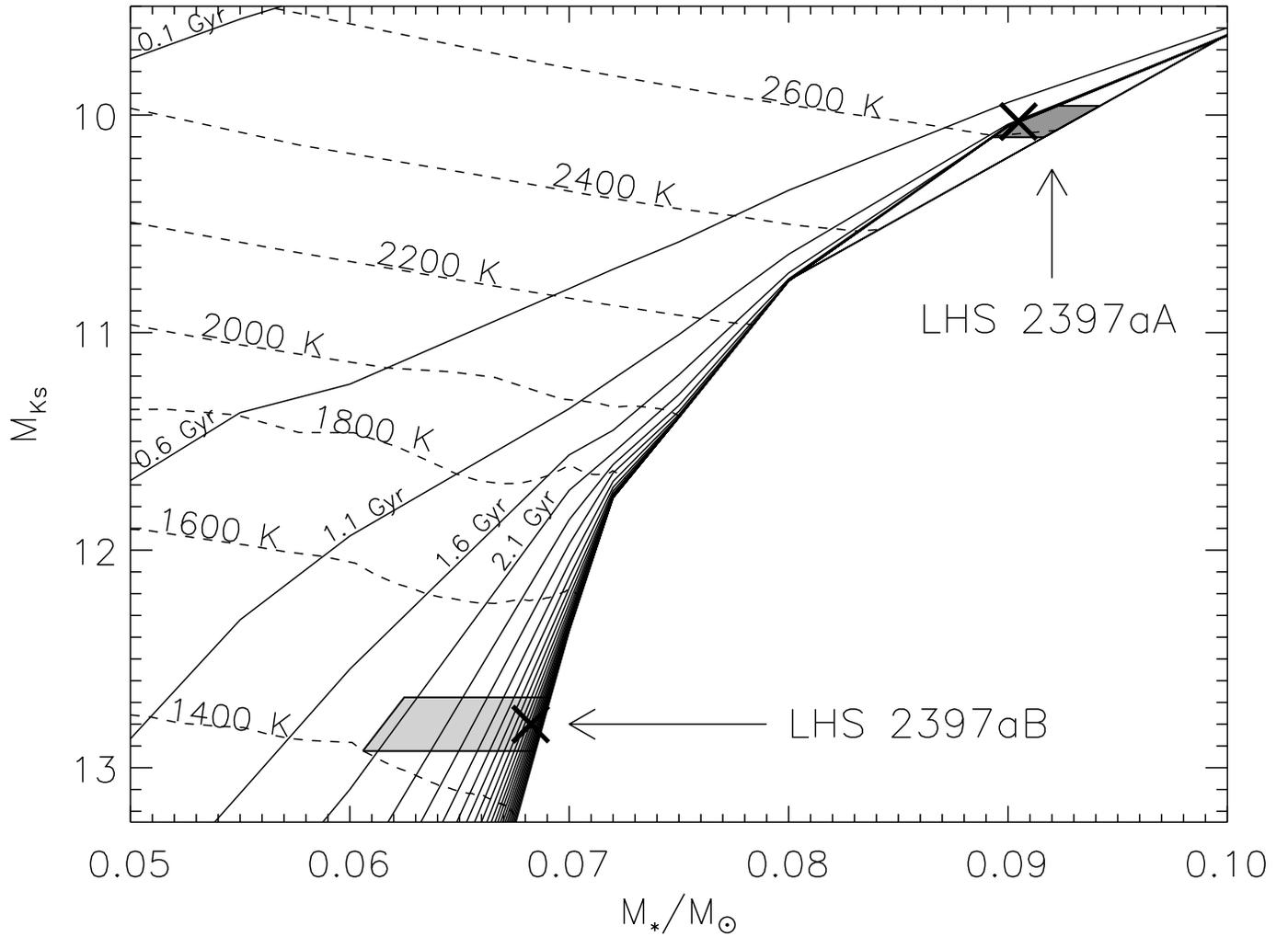}\caption{Figure 3 from \cite{fre02}. Note that the isochrones are different for this plot compared to all the other plots in this paper (but the isotherms are the same). More detail about LHS 2397a can be found in \cite{fre02}. LHS 2397aB (mass 0.068 $M_{\sun}$) is one of the tightest brown dwarf companions ever to be imaged around a star.\label{fig2c}}
\end{figure}

\begin{figure}
\includegraphics[angle=90,width=\columnwidth]{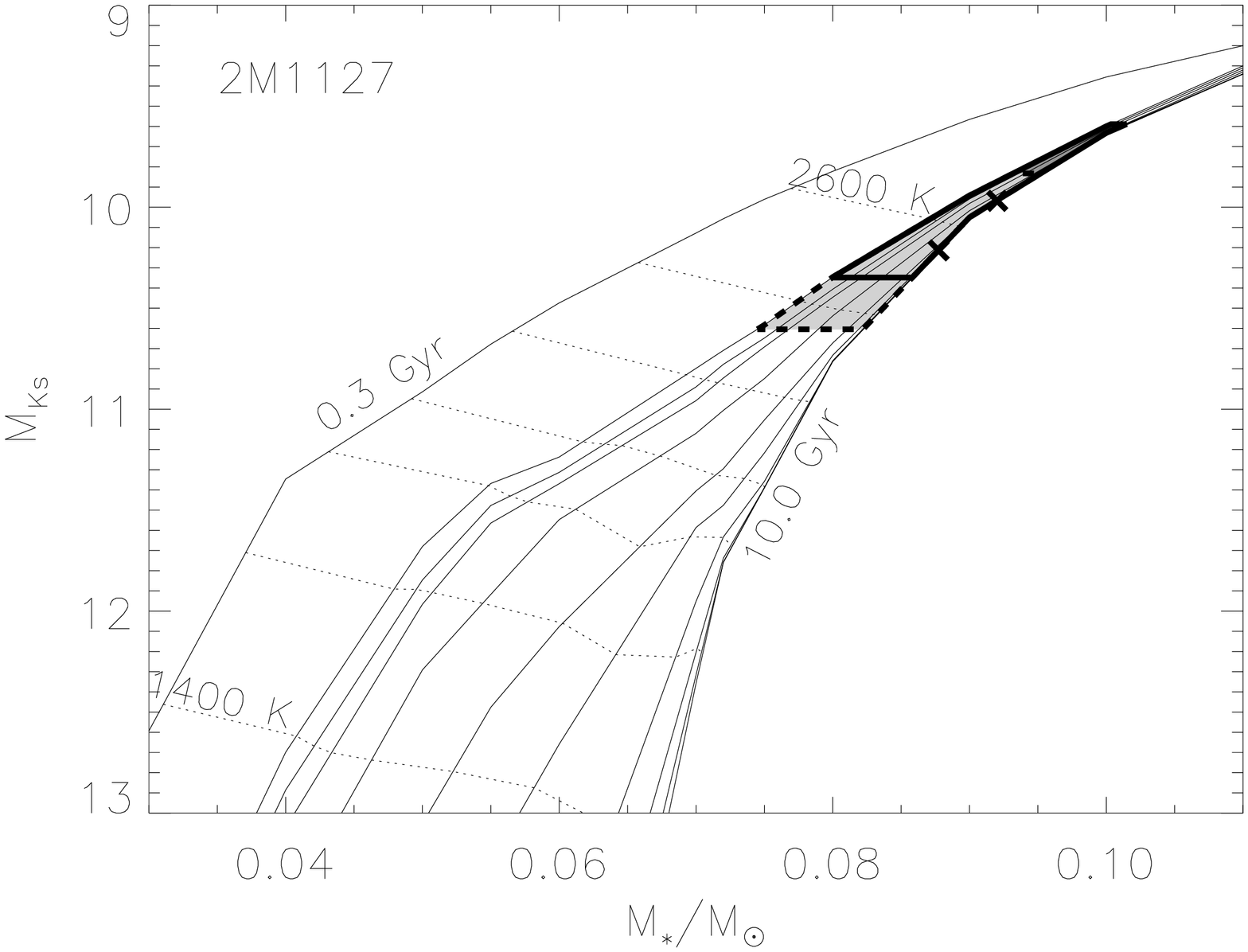}\caption{The same as for figure 3 except for 2M1127. Note the larger errors in $M_{Ks}$ compared to 2M0746 since we have to currently rely on a photometric parallax for this system. The
models suggest a primary mass of 0.092 $M_{\sun}$ with a range $
0.080-0.101 M_{\sun}$ with temperatures of 2650 K (2460-2810K). For
the secondary the models suggest a mass of $0.087 M_{\sun}$ with a
range $ 0.074-0.095 M_{\sun}$ with temperatures of 2540K (2330-2710
K). \label{fig2d}}
\end{figure}

\begin{figure}
\includegraphics[angle=90,width=\columnwidth]{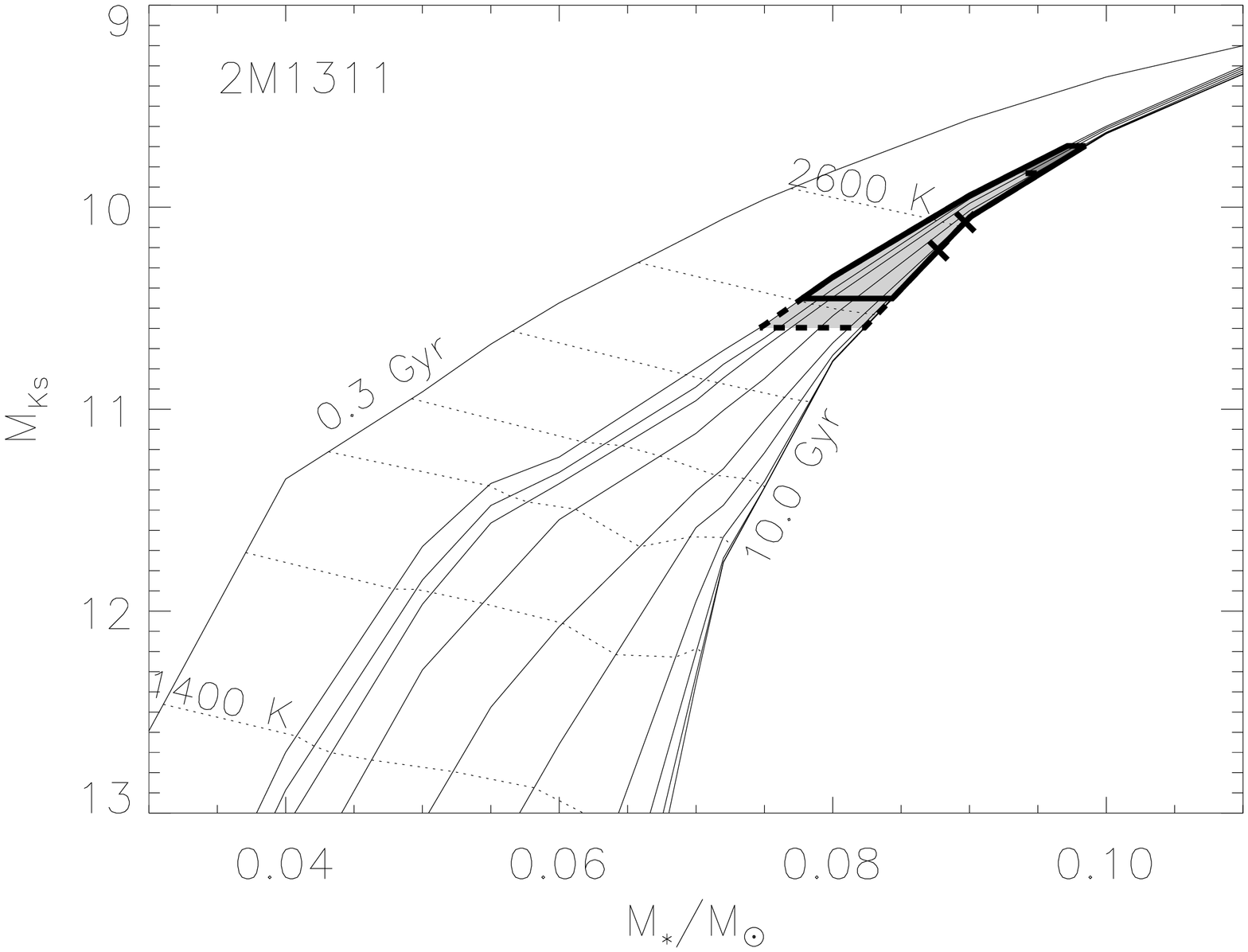}\caption{The same as for figure 3 except for 2M1311. Note the larger errors in $M_{Ks}$ compared to 2M0746 since we have to currently rely on a photometric parallax for this system. The
models suggest a primary mass of 0.089 $M_{\sun}$ with a range $
0.078-0.098 M_{\sun}$ with temperatures of 2610 K (2400-2760K). For
the secondary the models suggest a mass of $0.088 M_{\sun}$ with a
range $ 0.074-0.095 M_{\sun}$ with temperatures of 2540K (2331-2710
K). \label{fig2e}}
\end{figure}

\begin{figure}
\includegraphics[angle=90,width=\columnwidth]{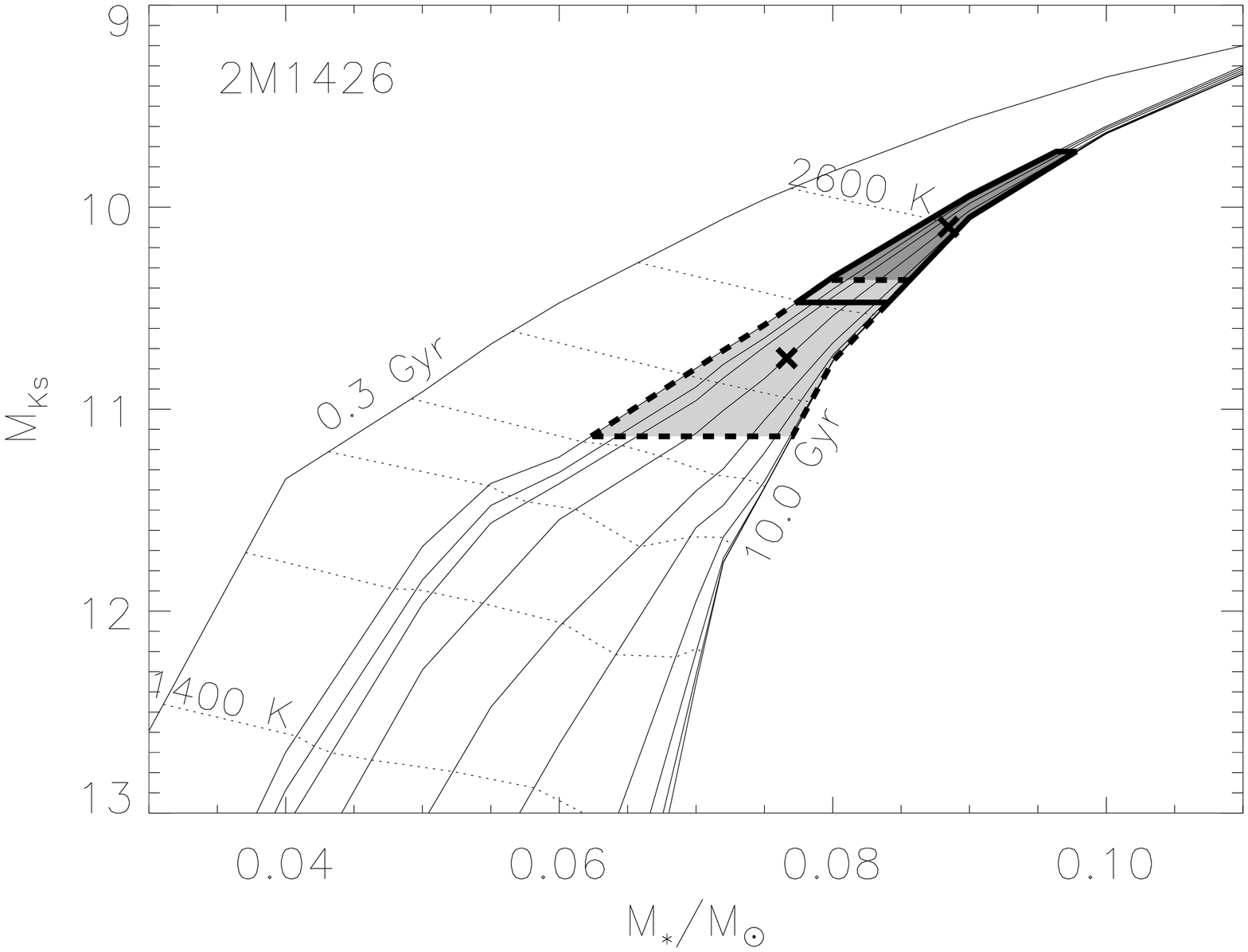}\caption{The same as for figure 3 except for 2M1426. Note the larger errors in $M_{Ks}$ compared to 2M0746 since we have to currently rely on a photometric parallax for this system.  The
models suggest a primary mass of 0.088 $M_{\sun}$ with a range $
0.077-0.097 M_{\sun}$ with temperatures of 2560 K (2400-2750K). For
the secondary the models suggest a mass of $0.076 M_{\sun}$ with a
range $ 0.062-0.085 M_{\sun}$ with temperatures of 2280K (2016-2480
K).\label{fig2f}}
\end{figure}

\begin{figure}
\includegraphics[angle=90,width=\columnwidth]{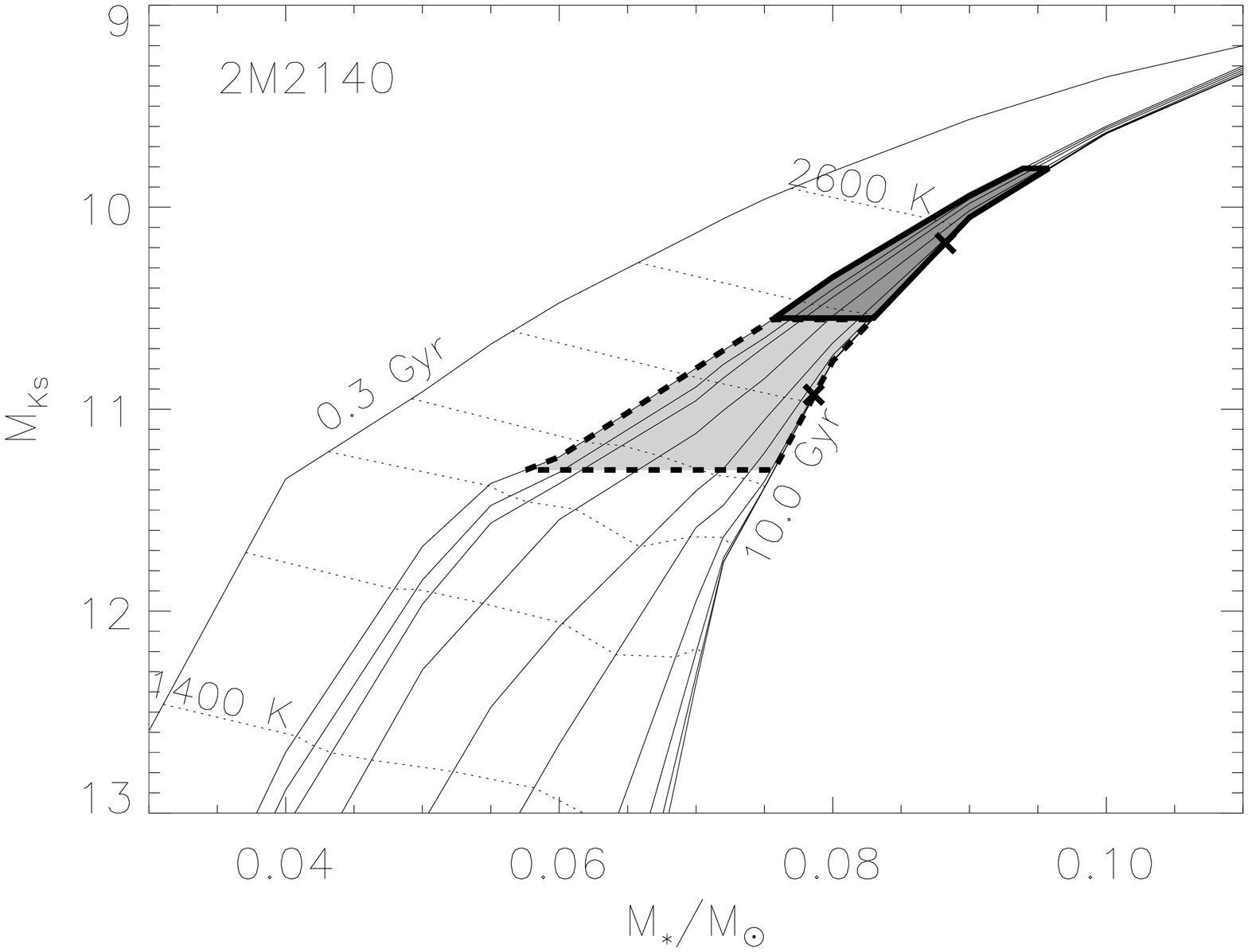}\caption{Same as for figure 3 except for 2M2140. Note the larger errors in $M_{Ks}$ compared to 2M0746 since we have to currently rely on a photometric parallax for this system.  The
models suggest a primary mass of 0.088 $M_{\sun}$ with a range $
0.075-0.095 M_{\sun}$ with temperatures of 2560 K (2360-2720K). For
the secondary the models suggest a mass of $0.078 M_{\sun}$ with a
range $ 0.057-0.083 M_{\sun}$ with temperatures of 2210K (1880-2390
K).\label{fig2g}}
\end{figure}

\begin{figure}
\includegraphics[angle=90,width=\columnwidth]{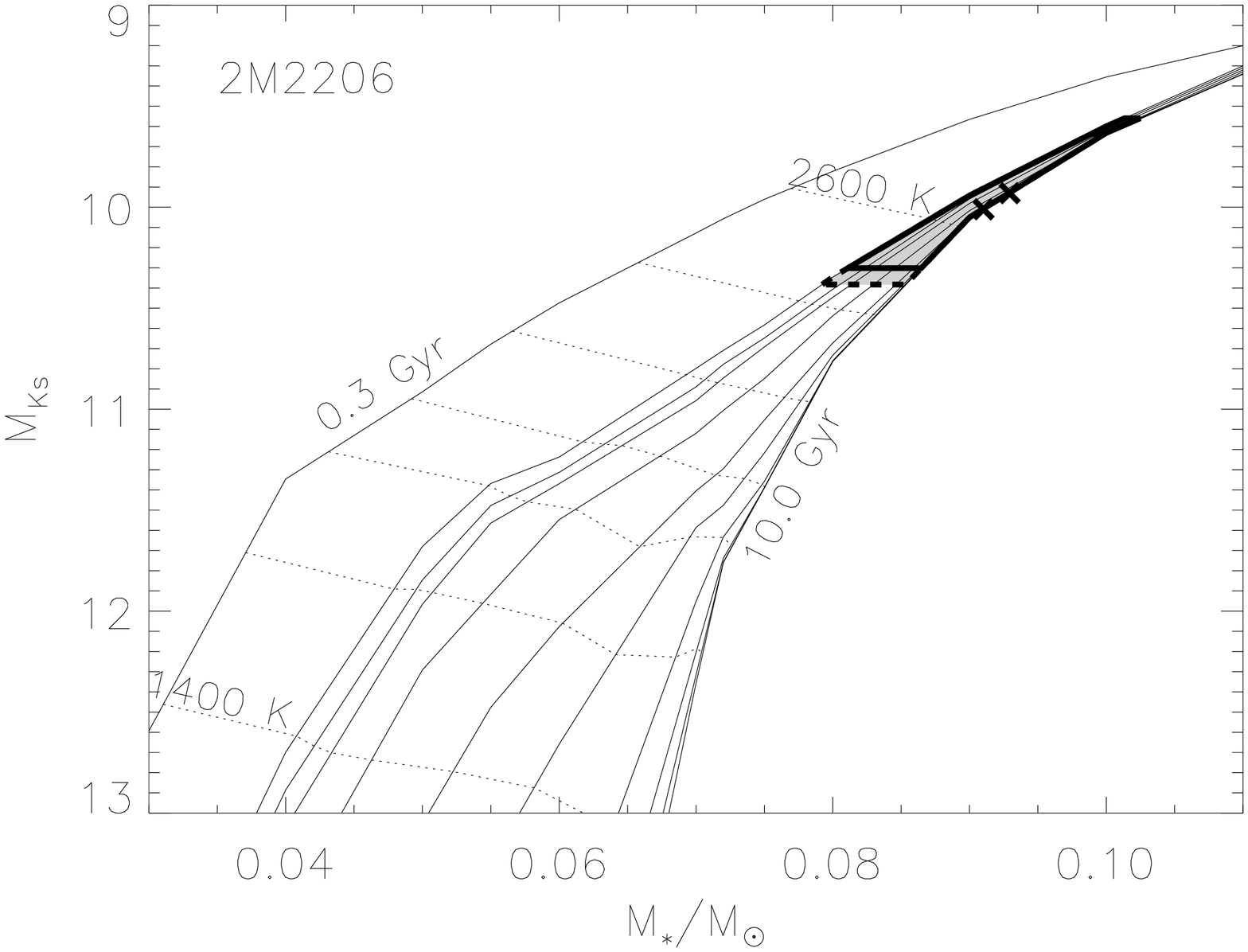}\caption{The same as for figure 3 except for 2M2206. Note the larger errors in $M_{Ks}$ compared to 2M0746 since we have to currently rely on a photometric parallax for this system. The
models suggest a primary mass of 0.092 $M_{\sun}$ with a range $
0.081-0.102 M_{\sun}$ with temperatures of 2670 K (2480-2830K). For
the secondary the models suggest a mass of $0.091 M_{\sun}$ with a
range $ 0.079-0.100 M_{\sun}$ with temperatures of 2640K (2440-2790
K). \label{fig2h}}
\end{figure}

\begin{figure}
 \includegraphics[angle=90,width=\columnwidth]{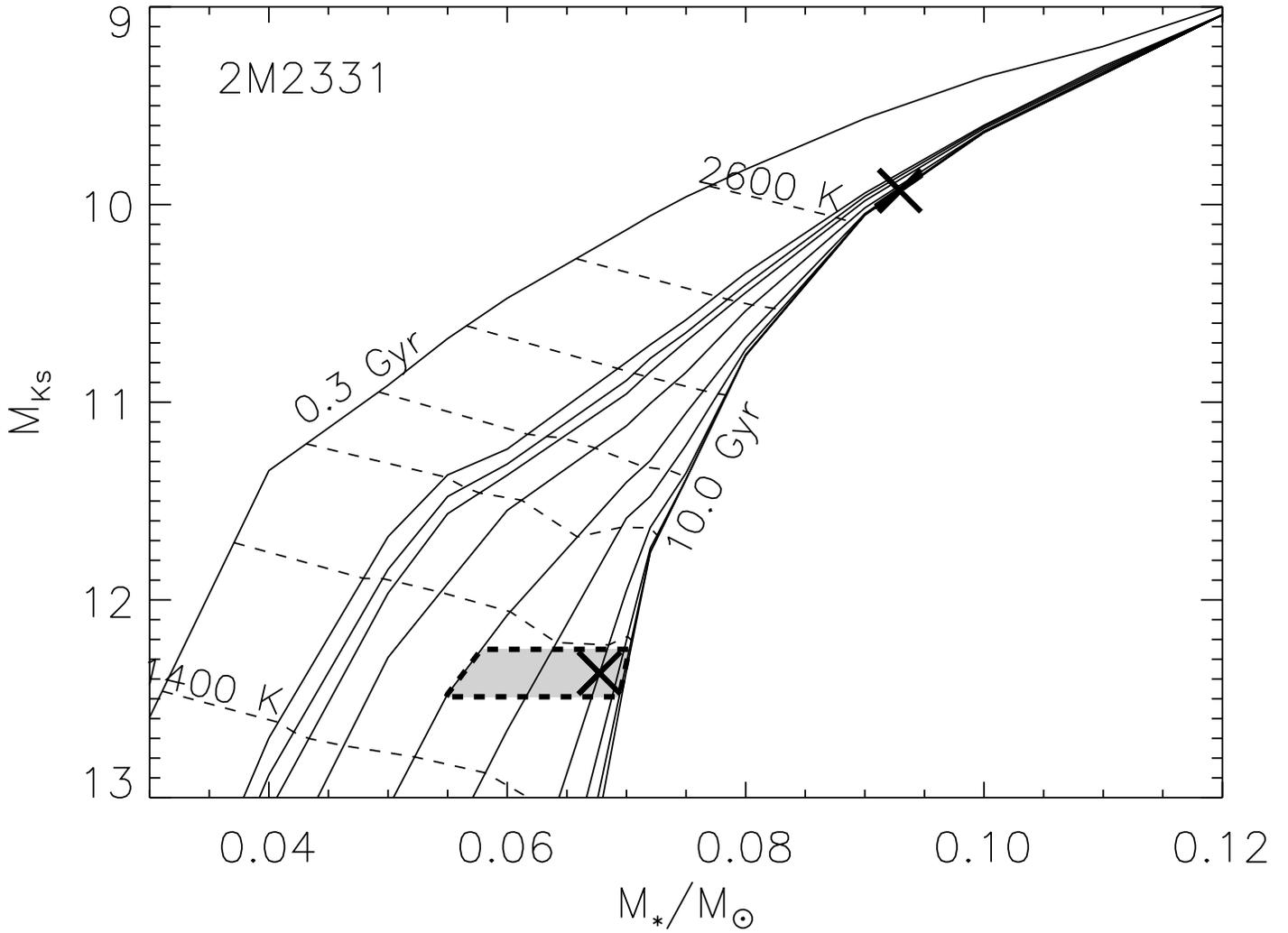}\caption{The
same as for figure 3 except for 2M2331. This system is associated with
the F8 star HD221356 \citep{giz00} which has a Hipparcos parallax of
$26.2\pm0.6$ pc. The system is very wide with a separation of 0.057 pc
between HD221356 and the 2M2331 system \citep{giz00}. The models suggest
a mass for 2M2331A of 0.093 $M_{\sun}$ with a range $ 0.091-0.095
M_{\sun}$ with temperatures of 2670 K (2640-2700K). For 2M2331B the
models suggest a mass of $0.067 M_{\sun}$ with a range $ 0.055-0.070
M_{\sun}$ with temperatures of 1560K (1470-1584 K), making 2M2331B a definite brown dwarf companion. \label{fig2i}}
\end{figure}

To estimate masses for these objects we will need to rely on
theoretical evolutionary tracks for VLM stars and brown
dwarfs. Calibrated theoretical evolutionary tracks are required for
objects in the temperature range 1400-2600 K. Recently such a
calibration has been performed by two groups using dynamical
measurements of the M8.5 Gl569B brown dwarf binary. From the dynamical
mass measurements of the Gl569B binary brown dwarf \citep{ken01,lan01} it was found that the \cite{cha00} and \cite{bur00}
evolutionary models were in reasonably good agreement with
observation.  In Figures
\ref{fig2} to \ref{fig2i} we plot the latest DUSTY
models from \cite{cha00} which have been specially integrated across
the Ks filter so as to allow a direct comparison to the
2MASS Ks photometry (this avoids the additional error of converting
from Ks to K for very red objects). We extrapolated the isochrones from 0.10 to 0.11 $M_{\sun}$ to cover the extreme upper limits of some of the primary masses in the figures.   
 
\placefigure{fig3} 
\placefigure{fig4}
\placefigure{fig5}
\placefigure{fig6}
\placefigure{fig7}
\placefigure{fig8}
\placefigure{fig9}
\placefigure{fig10}
\placefigure{fig11}

We estimate the masses of the components based on the age range of
0.6-7.5 Gyr and the range of $M_{Ks}$ values. The maximum mass relates to
the minimum $M_{Ks}$ and the maximum age of 7.5 Gyr. The minimum mass
relates to the maximum $M_{Ks}$ and the minimum age of 0.6 Gyr. These
masses are listed in Table
\ref{tbl-2} and illustrated in Figures \ref{fig2} to \ref{fig2i} as filled polygons. 

At the younger ages ($<1 Gyr$), the primaries may be on the
stellar/substellar boundary, but they are most likely VLM stars. The
substellar nature of the companion is very likely in the case of
2M2331B and LHS 2397aB, possible in the cases of 2M0746B, 2M1426B, and
2M2140B, and unlikely in the cases of 2M1047B, 2M1127B, 2M1311B, and
2M2206B which all appear to be VLM stars like their primaries. Hence
two of the companions are brown dwarfs, three others may also be substellar, and
four are likely VLM stars.

\section {Discussion}
\subsection {The binary frequency of M8.0-L0.5 stars}

We have carried out the largest flux limited ($Ks<12$) high spatial
resolution survey of M8.0-L0.5 primaries. Around these 39 M8.0-L0.5
targets we have detected 9 systems that have companions. Since our
survey is flux limited we need to correct for our bias toward
detecting equal magnitude binaries that ``leak'' into our sample from
further distances. For example, an equal magnitude M8 binary could
have an integrated 2MASS magnitude of Ks=12 mag but be actually
located at 36 pc whereas a single M8 star of Ks=12 would be located
just 26 pc distant. Hence our selection of $Ks<12$ leads to
incompleteness of single stars and low mass ratio ($q\tbond M_2/M_1$) binaries
past $D \sim 26$ pc. More exactly, 88\% of our binary systems are
within 29.1 pc (distances calculated using only the {\it primary's}
apparent magnitude) and 88\% of our single stars are within 23.4
pc. Therefore, we are probing $\sim(29.1/23.4)^3=1.92$ times more
volume with the brighter (combined light) of the binaries compared to
the single (hence fainter) M8.0-L0.5 stars. Hence, the corrected
binary frequency is $9/39/1.92 = 12\pm4 \%$ (where the error is only Poisson error).

There is another selection effect due to the instrumental
PSF which prevents detection of very faint companions very close to the
primaries. At the smallest separations of  $0.1-0.2\arcsec$ we are only sensitive to relatively bright companions of $\Delta K^\prime
\la 1 mag$. Much fainter companions
($\Delta K^\prime
\sim 5 mag$) can be detected at slightly wider ($\sim 0.25\arcsec$)
separations, and very low mass companions ($\Delta H \sim 10 mag$)
could be detected at $\sim 1\arcsec$ separations in our deep 720s H
images. Therefore, we are likely insensitive to faint ($\Delta K^\prime >
1.0$) companions in the separation range of $0.1-0.2\arcsec$. However,
if we assume that the mass ratio ($q$) distribution for M8.0-L0.5
stars is close to flat (as it is for M0-M4 binaries; \cite{fis92}), then we would expect at
least as many binaries with $\Delta K^\prime > 1.0 $ as $\Delta K^\prime < 1.0$
mag. Although we do not have enough data currently to
definitely derive the ${q}$ distribution for M8.0-L0.5 binaries, we can note that
for the four systems with separations $>0.2
\arcsec$ we observed an equal number of $\Delta K^\prime > 1.0 $ as $\Delta K^\prime
< 1.0$ mag systems. So it appears reasonable that there should also be
an equal number of $\Delta K^\prime > 1.0 $ as $\Delta K^\prime < 1.0$ mag systems
in the range $0.1-0.2 \arcsec$. Consequently, based on our detection of five
systems with $\Delta K^\prime < 1 $ mag with separations of $0.1-0.2\arcsec$
we would expect to have $\sim 5$ systems with $\Delta K^\prime > 1.0 $ in the
range $0.1-0.2\arcsec$. In reality we detected no systems with $\Delta
K^\prime > 1.0 $ with separations $0.1-0.2\arcsec$. Therefore, to correct for
instrumental insensitivity we need to increase the number of binary
systems by 5 in the range $0.1-0.2\arcsec$. Based on this assumption
about the mass ratio distribution there should be $\sim10$ binaries
from $0.1-0.2\arcsec$ when correcting for our instrumental
insensitivity. Therefore, the total count for all separations $>0.1
\arcsec$ should be $14\pm4$ systems assuming a Poisson
error. Therefore, the corrected M8.0-L0.5 binary frequency would be
$14/39/1.92 = 19\pm7\%$ for separations $> 0.10\arcsec$ or $\ga 2.6$ AU.
Hence we have a range of possible volume-limited binary frequencies (BF) from $12\%$ ($q\sim 1$) up to $19\%$ (where $q$ is assumed to be flat necessitating a large correction for insensitivity).

As a check we can re-derive these values of the true volume-limited BF in a manner after \cite{bur03} for the two limiting cases of $q=1$ up to a flat $q$ distribution. We can write,

\begin{equation}
{BF = -BF^{obs}/(BF^{obs}+\alpha(BF^{obs}-1))}
\end{equation}

 where $BF^{obs}$ is the observed uncorrected binarity ($BF^{obs}=9/39=23\%$), and where $\alpha$ is the fractional increase in volume sampled for binaries with a flux ratio $\rho = F_B/F_A$ and a flux ratio distribution $f(\rho)$. As in \citep{bur03} we can calculate $\alpha$ by,

\begin{equation}
{\alpha} \equiv \frac{\int_0^1{(1+{\rho})^{3/2}f({\rho})d{\rho}}}{\int_0^1{f({\rho})d{\rho}}}
\end{equation}

We consider two limiting cases for the $f(\rho)$
distribution: 1) if all the systems are equal magnitude ($q=\rho =1$)
then $\alpha=2^{(3/2)}=2.8$ and the BF=12\%; 2) if there is a flat
$f(\rho)$ distribution then $\alpha=1.9$ and the
BF=19\%. Consequently, the binary frequency range is
12-19\% which is identical to the range estimated above. Later we will see (Figure \ref{fig_q}) the true $f(\rho)$ is indeed a compromise between flat
and unity; hence we split the difference and adopt a binary frequency of $15\pm7\%$ where the error
is the Poisson error (5\%) added in quadrature to the ($\sim4\%$) uncertainty
due to the possible range of the $q$ distribution ($1.9<\alpha<2.8$). {\it It appears that for systems with separations $2.6<a<300$ AU the M8.0-L0.5 binary
frequency is within the range $15\pm7\%$}.

\placefigure{fig12}

\begin{figure}
\includegraphics[angle=0,width=\columnwidth]{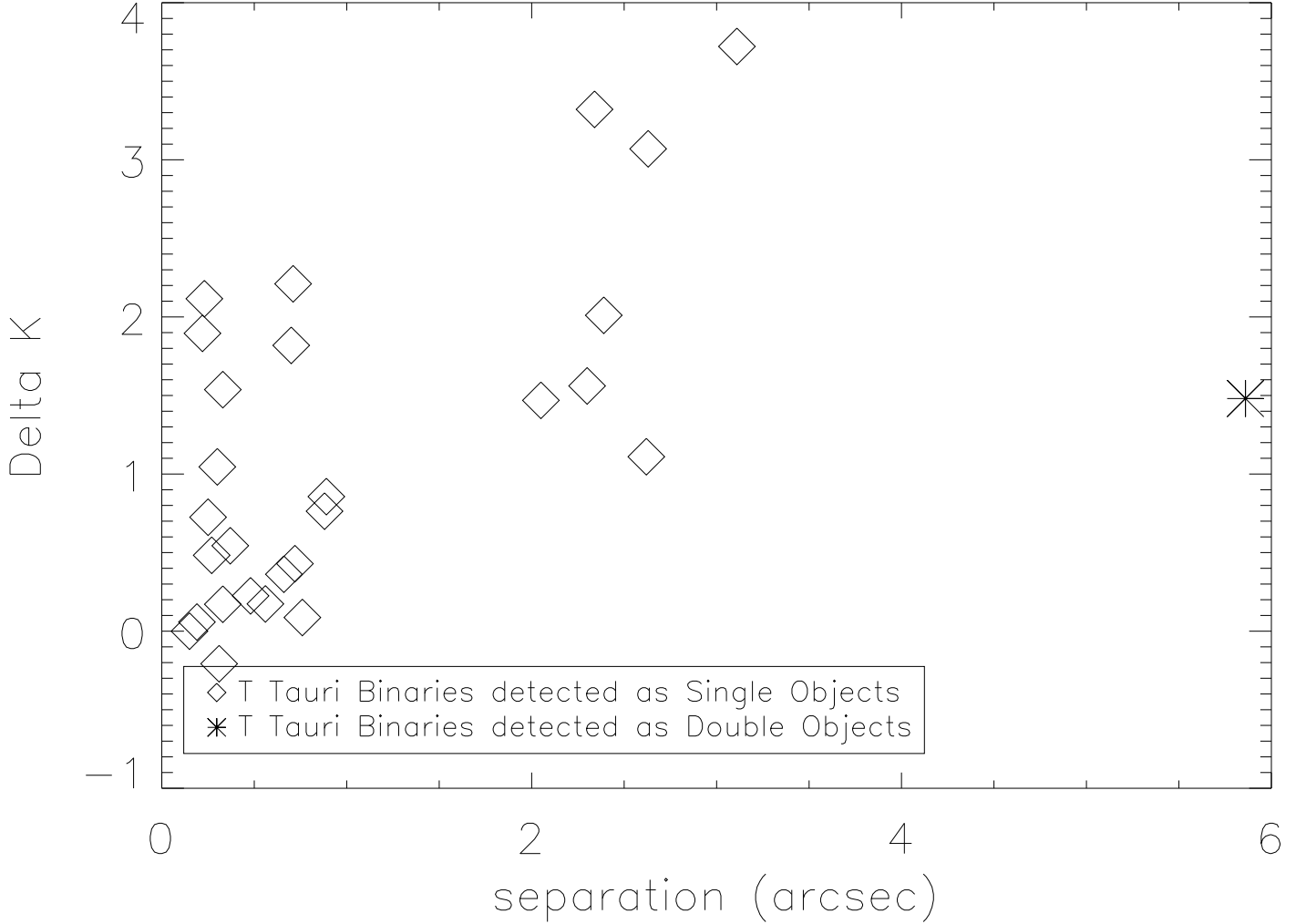}\caption{
It is possible that the reason none of these 2MASS based surveys have
detected wide low mass binaries is because the 2MASS pipeline flags
blended binaries as ``extended'' and they are left out of the point
source catalog completely. To check the validity of this possibility we plot above
all the T Tauri binaries from \cite{whi01} found in the 2MASS second
release point source catalog. None of these binaries were classified
as extended sources sources, hence they each had at least one entry in the
2MASS point source catalog. However, the low angular resolution of
2MASS resulted in only systems with separations greater than $\sim
4\arcsec$ actually being classified as binaries. All systems tighter
than $3\arcsec$ were unresolved and classified as single stars. A
fainter sample of white dwarf binaries yielded the same result. We
found no examples of binaries tighter than $2\arcsec$ being removed
from the point source catalog. Therefore, we feel the lack of
detection of wide low mass systems is not a selection effect from the
initial use of the 2MASS point source catalog to define candidate low
mass systems by \cite{giz00, kir00, bur03, cru02}.
\label{fig13b}}
\end{figure}

Our M8.0-L0.5 binary fraction range of
$15\pm7\%$ is marginally consistent with the $28\pm 9\%$ measured for more
massive M0-M4 dwarfs \citep{fis92} over the same separation/period range
($2.6<a<300$ AU) probed in this study. However,
\cite{fis92} found a binary fraction of $32\pm9\%$ over the whole
range of $a>2.6 AU$. If we assume that there are {\it no} missing low
mass wide binary systems with $a>300$ AU (this is a good assumption
since such wide $sep\ga 15\arcsec$ systems would have been easily detected in the
2MASS point source catalog as is illustrated in Figure \ref{fig13b}), then our binary fraction of $15\pm7\%$
would be valid for all $a>2.6$ AU and would therefore be slightly
lower than $32\pm9\%$ observed for M0-M4 dwarfs with $a>2.6$ AU by
\cite{fis92}. Hence it appears VLM binaries ($M_{tot}<0.185
M_{\sun}$) are less common (significant at the 95\% level) than M0-M4 binaries over the whole range
$a>2.6$ AU.

\subsection{The separation distribution function for M8.0-L0.5 binaries}

The M8.0-L0.5 binaries are much tighter than M0-M4 dwarfs in the
distribution of their semi-major axes. The M8.0-L0.5 binaries appear to
peak at separations $\sim4$ AU which is significantly tighter than the
broad $\sim30$ AU peak of both the G and M star binary distributions
\citep{duq91, fis92}. This cannot be a selection effect
since we are highly sensitive to all M8.0-L0.5 binaries with $sep>20-300$
AU (even those with $\Delta H>10$ mag). {\it Therefore, we 
conclude that M8.0-L0.5 stars likely have slightly lower binary
fractions than G and early M dwarfs, but have significantly smaller
semi-major axes on average}.

\section{The VLM binary population in general}

More observations of such systems will be required to see if these 
trends for M8.0-L0.5 binaries hold over bigger samples. It is interesting to note that in
\cite{rei01a} an HST/WFPC2 survey of 20 L stars found 4 binaries and a
similar binary frequency of 10-20\%. The widest L dwarf binary in
\cite{kor99} had a separation of only 9.2 AU. A smaller HST survey of 10 T
dwarfs by \cite{bur03} found two T binaries and a similar binary
frequency of $9^{+15}_{-4}\%$ with no systems wider than 5.2 AU. Therefore,
it appears all M8.0-L0.5, L, and T binaries may have similar binary
frequencies $\sim 9-15\%$ (for $a>3$ AU). 

\placetable{tbl-3}
\placefigure{fig13}

\begin{figure}
\includegraphics[angle=0,width=\columnwidth]{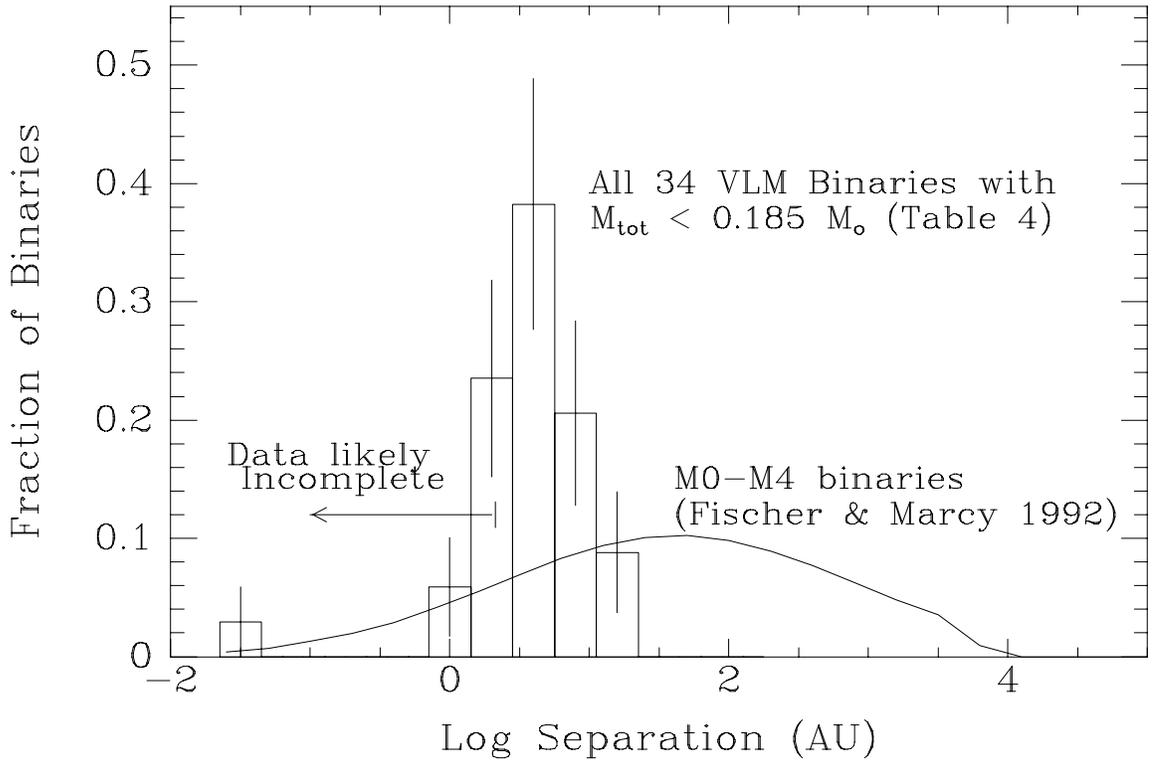}\caption{
Here we plot all 34 VLM binaries from Table \ref{tbl-3}. Note how VLM
binaries appear to have smaller separations compared to the M0-M4
binaries of \cite{fis92}. Both distributions are normalized to unity
binary fraction. We have not tried to correct for instrumental
incompleteness. Hence we underestimate the number of VLM binaries with
$a<2$ AU. Poisson error bars are plotted; the sharp peak at 4 AU and
the lack of any wide ($a>16 AU$) systems are real features of the
distribution and are significantly different from that observed in
more massive M0-M4 binaries ($M_{tot} \ga 0.3 M_{\sun}$). Binary systems with $M_{tot}<0.185
M_{\sun}$ appear $\sim10$ times tighter compared to just slightly ($2-5$ times) more
massive M0-M4 binaries. We have converted the semi-major axis
distribution values of \cite{fis92} to observed separation by dividing
the semi-major axis values by 1.26 (see equation 7 in \cite{fis92}).
\label{fig1c}}
\end{figure}

In Table \ref{tbl-3} we list all the currently known VLM binaries
(defined in this paper as $M_{tot}<0.185 M_{\sun}$) from the
high-resolution studies of
\cite{bas99,mar99,kor99,rei01a,lan01,pot02a,bur03,bou02}. As can be
seen from Figure \ref{fig1c} VLM binaries have a $\sim 4$ AU peak in
their separation distribution function with no systems wider than 15
AU. 

\placefigure{fig14}

\begin{figure}
\includegraphics[angle=0,width=\columnwidth]{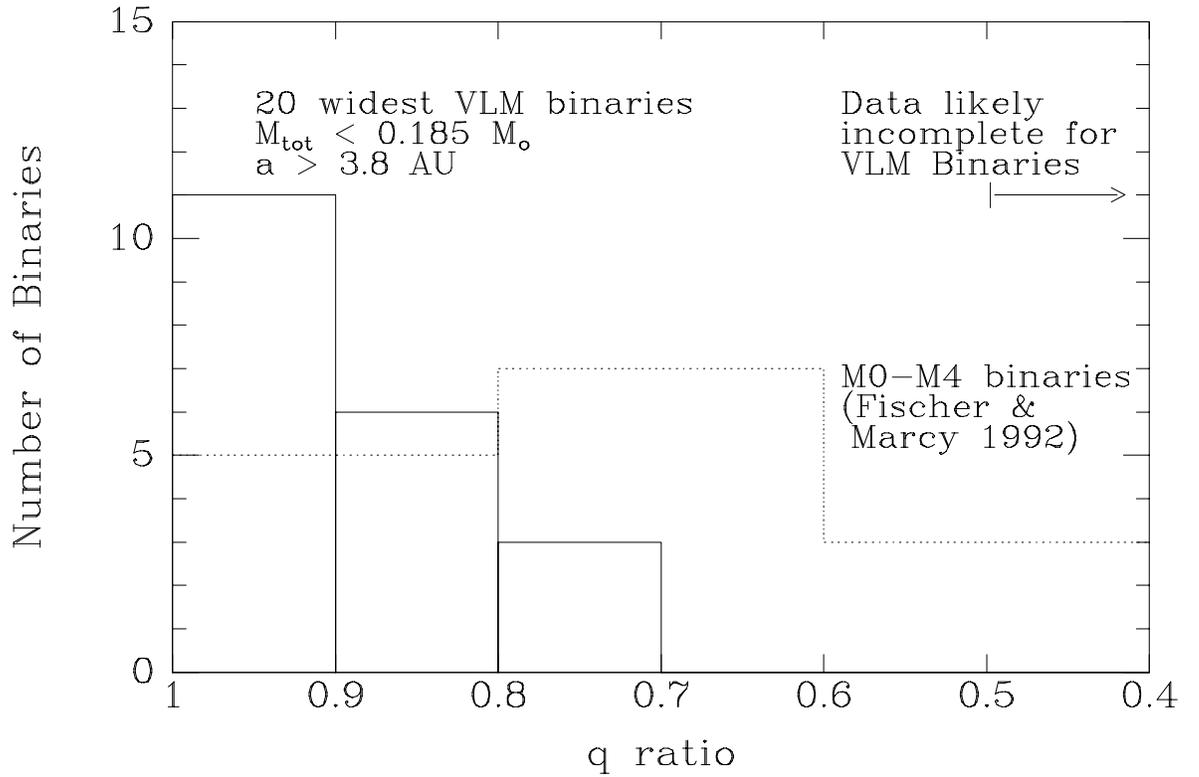}\caption{
Here we plot the 20 widest ($a>3.8$ AU) VLM binaries from Table
 \ref{tbl-3}. We see that VLM binaries appear to have a $q$ mass ratio peaked near
 unity. This is quite different from the much flatter distribution of
 M0-M4 binaries (dotted line) of \cite{fis92}. The difference is likely real, and not
 a insensitivity  effect, since the high-resolution AO and HST observations
 should be sensitive to mass ratios as low as $q\sim 0.5$ if $a\ga
 3.8$ AU \citep{bou02,bur03}. However, for very faint companions both HST/WFPC2 and AO surveys are likely incomplete for
 tight ($a\la 8$ AU) VLM binaries with $q\la 0.5$. Therefore, no conclusions can
 currently be drawn about the likelihood of VLM systems with $q<0.5$
 since we are largely insensitive to such systems in the current studies.
\label{fig_q}}
\end{figure}

From Figure \ref{fig_q} we see that most VLM binaries have nearly
equal mass companions, and no system has $q<0.7$. This VLM $q$
distribution is different from the nearly flat $q$ distribution of
M0-M4 stars \citep{fis92}. Since the HST surveys of \cite{mar99,rei01a,bur03,bou02} and our AO surveys
were sensitive to $1.0 >q\ga 0.5$ for systems with $a>4$ AU, the dearth
of systems with $0.8>q>0.5$ in Figure \ref{fig_q} is likely a real
characteristic of VLM binaries and not just a selection effect of
insensitivity. However, these surveys become insensitive to tight ($a<4$ AU) systems with $q<0.5$, hence the lack of detection of such systems may be purely due to insensitivity.

\subsection{Why are there no wide VLM binaries?}

It is curious that we were able to detect 8 systems in the range
$0.1-0.25\arcsec$ but no systems were detected past $0.5\arcsec$
($\sim 16 $ AU). This is surprising since we (as well as the HST surveys of \cite{mar99,rei01a,bur03,bou02}) are very sensitive to any
binary system with separations $>0.5\arcsec$ and yet none were
found. One may worry that this is just a selection effect in our target list from the spectroscopic surveys of \cite{giz00} and \cite{cru02}, since they only
selected objects in the 2MASS {\it point source catalog}. There is a possibility that such a catalog would select against $0.5\arcsec -
2.0\arcsec $ binaries if they appeared extended in the 2MASS
images. However, we found that marginally extended PSFs due to
unresolved binaries (separation $\la2 \arcsec $) were not being
classified as extended and therefore were not removed from the 2MASS
point source catalog. For example, Figure
\ref{fig13b} illustrates that no known T-Tauri binary from list of
\cite{whi01} was removed from the 2MASS point source
catalog. Although, due to the relatively poor resolution of 2MASS
(FWHM $\sim 2-3\arcsec$), only systems with separations $> 3\arcsec$
were classified as binaries by 2MASS, all the other T-Tauri binaries
were unresolved and mis-classified as single stars. In any case, we are
satisfied that no ``wide'' ($ 0.5\arcsec \la$ separation $\la 2\arcsec
$) VLM candidate systems were tagged as extended and removed from the
2MASS point source catalog. Therefore, the lack of a detection of any
system wider than $0.5\arcsec$ is not a selection effect of the
initial use of the 2MASS point source catalog for targets.

\placefigure{fig15}

Our observed dearth of wide systems is supported by the results of the HST surveys where out of 16 L 
and two T binaries, no system with a separation $>13$ AU was
detected. We find the widest M8.0-L0.5 binary is 16 AU while the widest L dwarf
binary is 13 AU \citep{bou02}, and the widest T dwarf binary is 5.2 AU
\citep{bur03}. However, M dwarf binaries just slightly more massive  ($\ga 0.2M_{\sun}$) in the field \citep{rei97a} and in the Hyades \citep{rei97b}
have much larger separations from 50-200 AU.

\begin{figure}
\includegraphics[angle=0,width=\columnwidth]{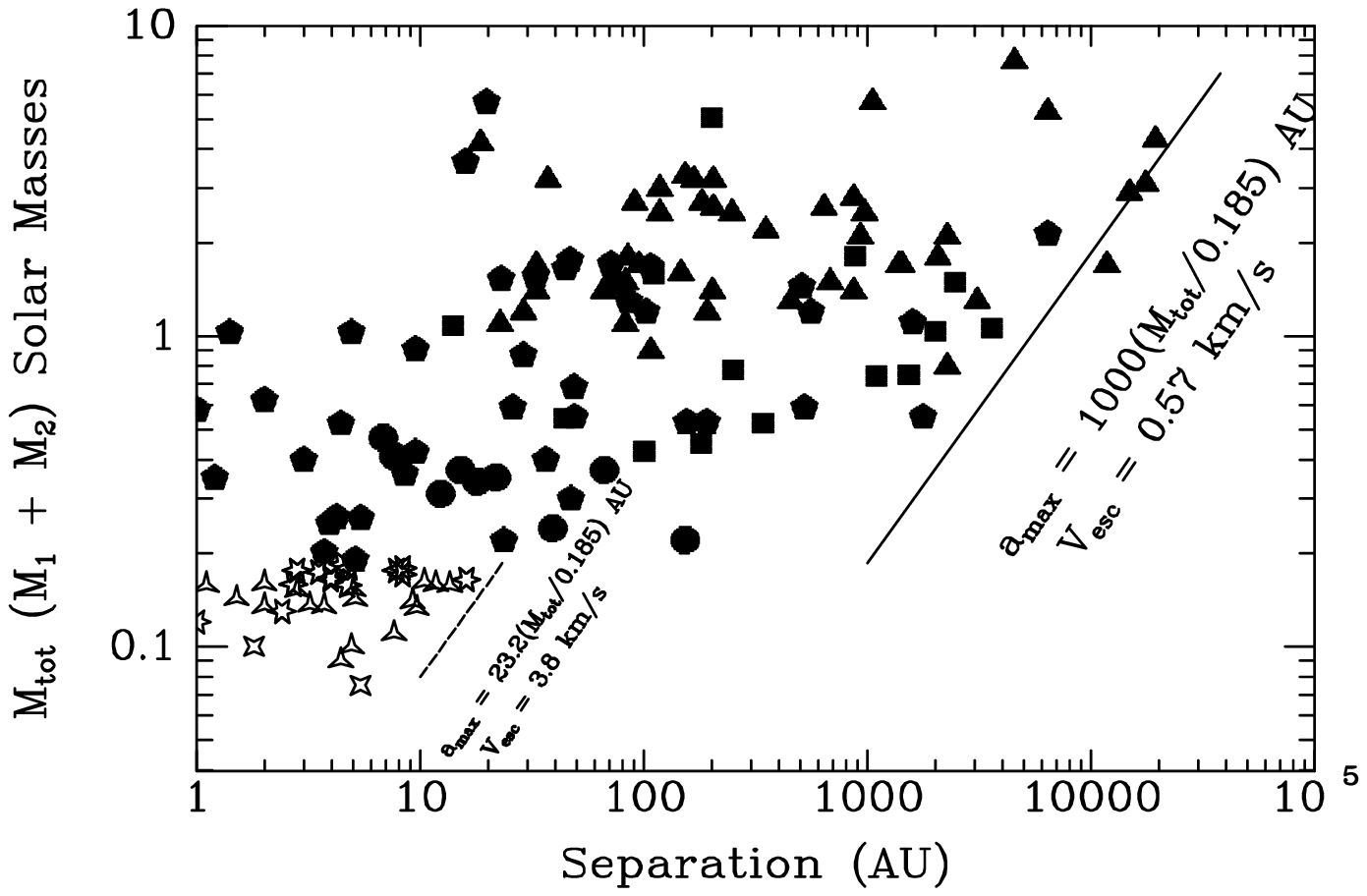}\caption{
Here we plot all 34 currently known VLM binaries. The nine
M8.0-L0.5 systems of this paper plus seven more VLM M binaries from Table
\ref{tbl-3} are plotted as open six sided stars. The L dwarf binaries of
\cite{kor99, mar99, rei01a, bou02} are shown as open triangles, and the two T dwarf binaries of
\cite{bur03} as open four-sided stars. For comparison, we have plotted
all the visual M0-A0 binaries inside 25pc from \cite{clo90} as solid
triangles. In addition, low mass Hyades binaries from \cite{rei97b}
are plotted as solid circles, low mass field M-dwarfs from
\cite{rei97a} are plotted as solid pentagons. All A0-M5 star/brown
dwarf systems are plotted as solid squares \citep{rei01a}. Note how
there are no low mass systems with separations $>16 AU$. It is seen that
for more massive binaries ($M_{tot}>0.185M_{\sun}$) the maximum observed
separation can be fit to $a_{max} = 1000(M_{tot}/0.185 M_{\sun})$ AU (solid
line). When $a=a_{max}$ the escape velocity is $V_{esc}=0.57$
km/s. However, for VLM binaries (where $M_{tot}<0.185 M_{\sun}$) the
best fit of the widest systems is $a_{max_{VLM}} =
23.2(M_{tot}/0.185 M_{\sun})$ AU (dashed line). For VLM systems with
$a=a_{max_{VLM}}$ the escape velocity is 3.8 km/s. {\it Hence VLM binaries
appear sharply tighter ($a<16$ AU) and have escape velocities at least 3
km/s higher than more massive wide binaries.}
\label{fig3}}
\end{figure}

\begin{figure}
 \includegraphics[angle=0,width=\columnwidth]{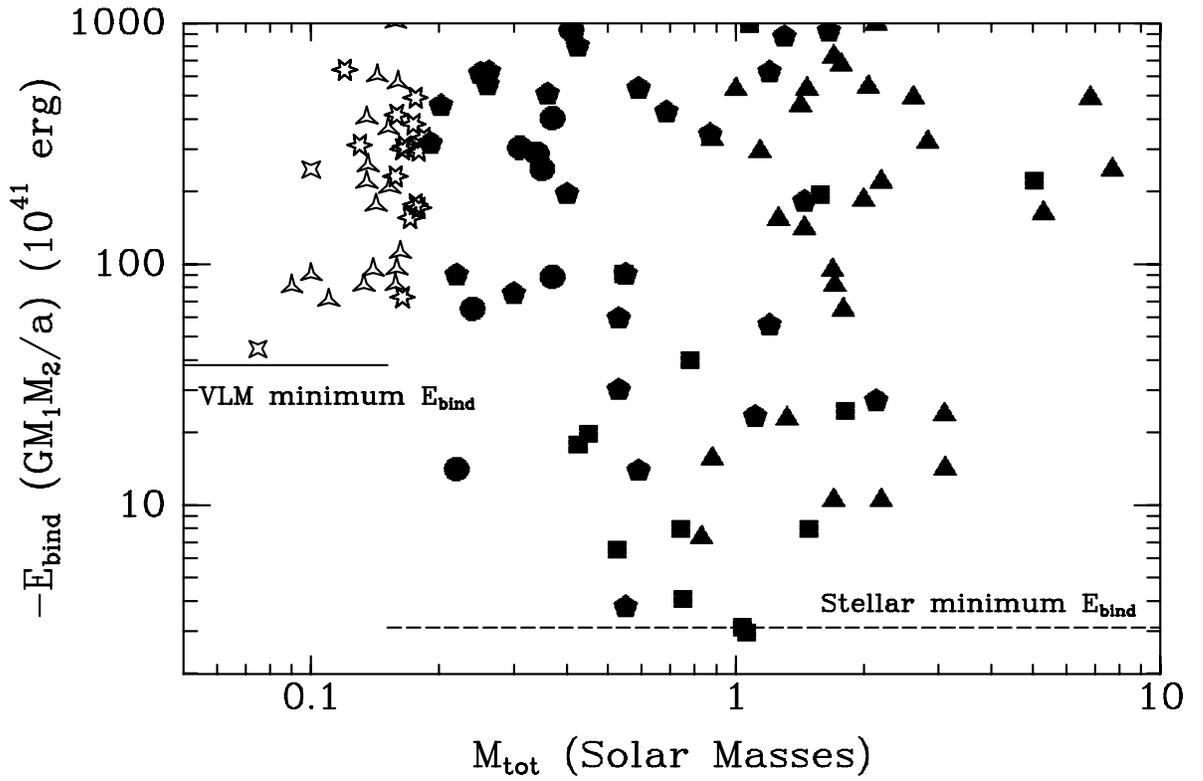}\caption{To
 see if low mass binaries really are more bound (i.e. ``harder'') we
 plot here the binding energy for all the systems in Figure
 \ref{fig3}. The data (individual masses and separations) are from Table \ref{tbl-3} and the symbols are the same as in Figure
 \ref{fig3}. We see that the widest VLM binaries (open symbols)
 are $\sim16$ times harder than their more massive wide binary
 counterparts (solid symbols). This effect was also noted by
 \cite{bur02}. \label{fig4}} \end{figure}

In Figure \ref{fig3} we plot the sum of primary and secondary
component masses as a function of the binary separation for all
currently known VLM binaries listed in Table \ref{tbl-3}. It
appears that all VLM and brown dwarf binaries (open symbols) are much
tighter than the slightly more massive M0-M4 binaries (solid symbols).

If we examine more massive (SpT=A0-M5) wide binary systems from 
\cite{clo90}, we see such wide binaries have maximum separations best fit by $a_{max} \sim
1000(M_{tot}/0.185 M_{\sun})$ AU in Figure \ref{fig3}. Any system with such a separation ($a =
a_{max}$) will be disrupted by a differential velocity impulse (or
``escape kick'') to one component of $V_{esc}> 0.57$ km/s (see the solid
upper line in Figure \ref{fig3}). However, if we try to fit the
VLM/brown dwarf binaries (defined here as systems with $M_{tot} <
0.185 M_{\sun}$) we find the maximum separation is better predicted by
$a_{max_{VLM}} \sim23.2(M_{tot}/0.185 M_{\sun})$ AU (lower dashed line in Figure
\ref{fig3}). Since these are much smaller separations we find that any
system with such a separation ($a =
a_{max_{VLM}}$)  will require a larger escape velocity kick of
$V_{esc}> 3.8$ km/s to become unbound.

%Hence, it appears that binaries with $M_{tot} < 0.185 M_{\sun}$ may
%have been subjected to an additional differential $ \sim3$ km/s
%kick which has dissolved all systems to the right of the lower dashed
%line (which bound all VLM systems) with $V_{esc}=3.8$ km/s in Figure
%\ref{fig3}. It is interesting to note that a kick of this order of
%magnitude was predicted (independently) by \cite{ste98,rep02} for
%brown dwarf formation by embryo ejection.

\placefigure{fig16}

To try and glean if VLM/brown dwarf binaries are really more tightly
bound, we compare their binding energy to that of more massive
binaries. Figure \ref{fig4} illustrates how the minimum binding energy
of binaries with $M_{tot} < 0.185 M_{\sun}$ is 16x harder than 
more massive field M-G binaries. In other words the widest binaries
with $M_{tot} < 0.185 M_{\sun}$ appear to be 16 times ``more bound'' than the
widest binaries with $M_{tot} > 0.185 M_{\sun}$. 

This hardening could
be a relic from dissipative star/disk interactions with the accretion disks
around each star \citep{mcd95} and/or it could be from an ejection event, dynamical decay, or fragmentation. We will briefly explore some of these possibilities in the next sections.

\subsubsection {Can ejection explain the lack of wide VLM binaries?}

\begin{figure}
 \includegraphics[angle=0,width=6in]{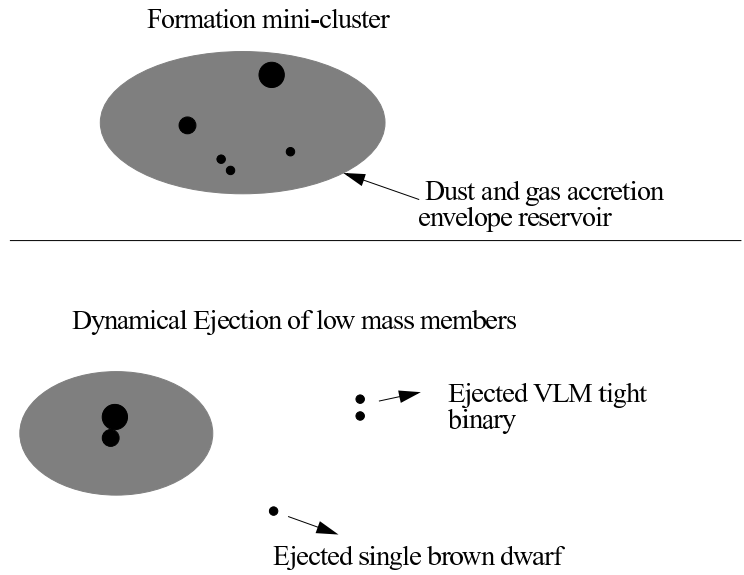}\caption{ A
simple cartoon of the hypothesized ejection of the low mass members of
a formation mini-cluster. Typically a dissolving mini-cluster will
evolve to a final hardened binary containing the most massive stars in
the cluster. The other lower mass members are ejected through close
encounters with the more massive members. If it is possible to eject
VLM binaries, only tightly bound VLM binaries could survive the
ejection process \citep{rep02}.
\label{cartoon}} \end{figure}

 \cite{rep02} suggest that VLM binary systems may have been ejected
 from their ``mini-cluster'' stellar nurseries in close triple or quadruple
 encounters with more massive objects early in their
 lives. Consequently, these ejected VLM objects are starved of
 accretion material and their growth is truncated (see a cartoon of
 this scenario in Fig. \ref{cartoon}).  \cite{ste98} estimate that around 5\% of ejecta from pentuple systems are 
binaries. The typical ejection velocity
 estimated by
\cite{ste98,rep02} is $\sim3$ km/s for single objects (on average these ejected single objects have masses$>0.185 M_{\sun}$). Hence one might hypothesize tight VLM binaries of similar total mass may be ejected at similar velocities; however, we caution that more detailed simulations are required to estimate realistic ejection velocities. In any case, only tightly bound binaries will survive the ejection process. This may explain the
additional tightness of the VLM/brown dwarf binaries in Figure
 \ref{fig3}. More loosely bound VLM binaries were dissolved
 (``kicked'' past the $V_{esc}=3.8$ km/s line in Fig.\ref{fig3}) when they
 were ejected from their mini-cluster. Hence, only relatively hard
 $a<16$ AU ($V_{esc}>3.8$ km/s) low mass binaries have survived until
 today.

%% Indeed, observations of lowest mass members of the Hyades
%% have $\sigma_{v} \sim 3$ km/s whereas the higher mass members have
%%$\sigma_{v} \sim 0.3$ km/s \citep{rei97b?}, which is observational
%% evidence that low mass objects have $\sim 3$km/s higher velocity
%% dispersions. 

\placefigure{fig17}

However, the ejection paradigm of \cite{rep02} as simulated in detail
 by \cite{bat02} only predict a VLM/brown dwarf binary fraction of $\la 5\%$ 
 which is below the $15\pm7\%$ observed for M8.0-L0.5 binaries and the $9^{+15}_{-4}$  observed for T dwarf binaries
 \citep{bur03}. Therefore, further simulations will be
 required to see if a larger ($\ga 9 \%$) binary frequency can be produced
 when low mass binaries are ejected from the mini-cluster.

\subsubsection {Is the dearth of wide low mass binaries due to Galactic dynamical evolution?}

%We see strong evidence that these low mass binary systems are
%systemactically and significantly harder than their more massive
%counterparts. This could imply that low mass binaries are ejected from
%thier birplaces by encouters with more massive objects in their ``mini-cluster''. However, it is also possible that low mass stars are
%prefferientially given more of a velocity kick in $\it{any}$ steallar
%encounter during the life of the binary. Hence, in the case of 2 low
%mass stars/brown dwarfs there will be always a greater likelihood of
%disruption at every stellar encounter compared to a binary composed of
%more massive stars. The impulse energy delivered to a binary from an
%encounter where the impact parameter (b) is much larger than the
%semimajor axis (a) is $\DeltaE = 7/3(GM_{p}a^2)/(V_{rel}b^3)$ where
%$V_{rel}$ is the relative velocity of the perturbing star of mass
%$M_{p}$ \cite{wei87}. 
%A catastrophic encounter leading to disruption
%of the binary will occur at a rate proportional to $a^(2/3)/M^(1/2)$
%hence low mass systems with only 0.1 the mass will suffer catastrophic
%encounters 3 times as often. 

Over the lifetime of a binary there will be many stochastic
encounters, which will increase the potential energy (and therefore
its separation) of the binary slowly over time. Eventually, these
encounters may also disrupt the binary. In fact, this disruption
timescale is roughly proportional to $M_{tot}/a$ for separations $<200$ AU
according to the detailed models of \cite{wei87}. As we can see from the solid line in 
Figure \ref{fig3}, a line of constant $M_{tot}/a$ or constant $V_{esc}$ (where $V_{esc}=0.57$
km/s) can fit the widest A0-M0  binaries, but the same line cannot fit the widest of the much tighter VLM  
binaries (open symbols). This is quite puzzling since wide ($>200$ AU)
VLM binaries should get wider throughout their lifetime 
until they reach $a=a_{max}\sim 1000(M_{tot}/0.185 M_{\sun})$ AU. By this point,
they have reached the average minimum binding energy
($-E_{bind_{min}}\sim3$x$10^{41}$ erg) of field wide binaries and are likely to be disrupted by a differential kick of only 0.6 km/s.

So why do we not observe wider VLM binaries if they should dynamically evolve to  $a=a_{max}\sim 1000(M_{tot}/0.185 M_{\sun}) $AU $\sim 1000 $AU over time?
It is critical to note that only binaries that are wider than
$1000(M_{tot}/M_{\sun})$ AU are wide enough to significantly evolve in the
galactic disk according the detailed models of
\cite{wei87}. Therefore, only binaries of initial separation ($a_{o}$) greater
than $a_{o}\sim 185$ AU for $M_{tot}=0.185 M_{\sun}$ will evolve to
the minimum binding energy (which corresponds to separation of
$a_{max} \sim 1000$ AU) over a 12 Gyr
lifetime. Any systems formed tighter than $1000(M_{tot}/M_{\sun})$ AU will
not dynamically evolve to significantly wider separations over time.

So why don't we detect any 185-1000 AU VLM binaries? The answer may be
very simple. If the formation process that forms VLM/brown dwarf
binaries cannot produce binaries with $a_{o}>1000(M_{tot}/M_{\sun})$ AU
then there will be no significant evolution of $a$ during the lifetime
of the binaries. In other words the observed $a$ distribution will be
similar to the initial distribution ($a_{o}$). Since we currently
observe no VLM systems with $a>100$ AU, the observed separation distribution for VLM binaries is likely the same as their initial distribution. Figures
\ref{fig3} and
\ref{fig4} suggest that $a_{o}$ for VLM stars is strongly truncated at
$\sim16$ AU.
In addition, the
observed values of $V_{esc}>3.8$ km/s and $-E_{bind}>50\times 10^{41}$ erg
are therefore likely relics of the formation of these VLM binaries
and need to be explained by any successful model of star/brown dwarf
formation.

\subsubsection{Could VLM binaries be the decay product of a dissolving mini-cluster?}

In the dynamical decay of these ``mini-clusters'' it is
likely that the most massive components become a hardened binary as
the lower mass objects are ejected by triple encounters (see bottom of
Figure \ref{cartoon}). \cite{rep02} have hypothesized that
VLM binaries may also be ejected from these mini-clusters since they
have very low masses and are tight enough to remain bound after
ejection. However, it is also possible that VLM binaries are not
ejected but instead persevere until they are the final most tightly bound binary remaining after the
``mini-cluster'' decays.

The decay of 3,4, and 5-body ``mini-clusters'' has been studied by
\cite{ste98}. The final residual binaries produced by such decays have
characteristics similar to those observed for VLM binaries. In
particular, \cite{ste98} predict that binaries composed of stars with
masses from $0.1-0.2 M_{\sun}$ would have separations of $\sim 10$ AU
compared to $\sim 30$ AU for more massive stars. Furthermore, they predict
the mass ratio $q$ to be in the range 0.6-1.0. Moreover, they predict
a fairly sharp cut-off at $\sim 10$ AU for ``wide'' low mass
binaries. All these predictions are roughly consistent with our
observations of the VLM binary population.

Since the final binary produced is typically biased towards the two or three
most massive members of the mini-cluster (true of $\sim99-98\% $
cluster decays; \cite{ste98}), the likelihood of the most massive
objects in the mini-cluster both having masses less than $ 0.090
M_{\sun}$ is rather small. Indeed, the binary fraction of $0.2<M_{tot}<0.4
M_{\sun}$ binaries predicted by \cite{ste98} is very small ($\le
1\%$) if each member of the cluster are picked randomly from the IMF. Even smaller binary fractions would be predicted for even lower mass
VLM binaries ($M_{tot}<0.185 M_{\sun}$). Therefore, only a fraction of
our VLM binaries (observed to have a binary frequency of $\sim 9-15\%$) are
likely the residual hardened binaries remaining after a mini-cluster
dissolves. 

Recently, \cite{dur01} have modeled a ``two-step IMF'' to produce a
more realistic binary population from F to early M spectral types. As
in \cite{ste98} the VLM binary characteristics predicted in general
are similar to those observed. In particular, this is true for their
models which include brown dwarfs in the IMF. However, they note that
models which include brown dwarfs overestimate the number of G-dwarf
binaries ($BF(1.0 M_{\sun})
\sim75\%$). Moreover, while these models do predict some binary VLM
systems ($BF(0.1 M_{\sun}\la 5\%$) they still underestimate the
frequency of such systems compared to the $15\pm7\%$ observed here for binaries in this mass range. 

%It may be more probable that a VLM binary is ejected from the cluster since
%it is typically of much lower mass than the other members of the
%cluster. Indeed, the simulation of \cite{bat02} shows the creation of
%an tight brown dwarf binary with a 6 AU separation. This binary is
%likely low enough in mass to be ejected from the mini-cluster but
%tight enough to stay bound during and after the ejection
%\citep{bat02}. However,
%\cite{bat02} caution that their simulation has only $\la 5 \%$ of
%brown dwarfs in the form of binaries. This is significantly less than
%the $9^{+15}_{-4}\%$ brown dwarf binarity observed by \cite{bur03} and the $15\pm 7\%$ observed here for M8-L0.5 binaries.

\subsubsection{Can we explain the truncation at $\sim16$ AU simply by an initial semi-major axis distribution produced by fragmentation?}

%If we simply consider a low mass fragmenting core which fragments into
%a VLM/brown dwarf binary we would expect the collapsing core mass to
%be $M_{core}\sim0.1 M_{\sun}$ (collapsing core masses as small as
%$\sim0.01 M_{\sun}$ are thought to be common if there is little
%angular momentum in the system). To collapse low mass cores with only
%$\sim 0.1 M_{\sun}$ will require less angular momentum and therefore
%we may have $R_{core}$ vary as $M$. 

One could avoid the problem of the unlikelihood of VLM binaries
surviving the dynamics of an ejection or cluster decay if one simply
produces VLM binaries in the same fashion that more massive binaries
are thought to form. This is commonly thought to be through the fragmentation
of molecular cloud cores. A fragmenting very low mass molecular cloud
core could directly produce a VLM binary without evoking any ejection
or decay processes.

It would be interesting to see if we can approximate a VLM binary
semi-major axis distribution produced by fragmentation by scaling
distributions of more massive binaries. If we assume $M_{tot} / a_{o}$
is roughly constant for wide binaries (see Figure \ref{fig3}), then we
expect the initial $a_{o}$ distribution to be somewhat tighter for VLM
binaries compared to more massive binaries. We can estimate the
$a_{o}$ for solar mass binaries from the well known young T-Tauri star
$a$ distributions. Young T-Tauri binaries are strongly suspected to
have formed by fragmentation
\citep{whi01}. We can utilize a HST sample of 29 T-Tauri binary systems
in Taurus to produce an initial separation distribution. Such a
distribution from the sample of \cite{whi01} has an $a_{o}$ peak at
$\sim30$ AU (as also found by \cite{lei93,ghe93}). We can scale this to match our observed VLM binary peak
of $\sim 4$ AU by simply scaling the distribution by $a_{o_{VLM}} \sim
0.13(a_{o_{T Tau}})$. Similarly we can scale the masses of the T Tauri
binaries to also match our mean VLM binary mass of $\sim 0.15
M_{\sun}$ by multiplying by the same factor of 0.13. In this manner we
have created a plausible fragmentation-produced VLM $a_{o}$
distribution. However, we have assumed that the other properties of the
distribution (such as the FWHM and tail distributions) can be scaled
along with the mean $a_{o}$ and $M_{tot}$. Assuming such a scaling is
valid globally, we see that this fragmentation $a_{o_{VLM}}$ distribution
has $\sim26\%$ of systems wider than $\sim40$ AU. Hence if our
observed VLM distribution is to match this scaled fragmentation
$a_{o}$ distribution, we would expect $\sim9$ of the 34 VLM/brown dwarf systems
observed (see Table
\ref{tbl-3}) to have separations greater than 40 AU. However, from
Figure \ref{fig3} { \it no VLM systems are observed to have
 separations greater than 16 AU}. Therefore, it appears very hard to
 explain the total lack of systems with separations greater than 16 AU
 by scaling the observed $a_{o}$ distribution of T-Tauri stars. In
 other words, if we scale the distribution to match the VLM $\sim4$ AU
 peak we over-produce wide systems (26\% of systems wider than 40 AU)
 compared to observations.  It may be that our ``toy model'' of simply
 scaling the T Tauri $a_{o}$ distribution is too simple an approach,
 but, it is a first step which shows that such a sharp cut-off
 ($a_{o}<16$ AU) is not easily produced by a 
 fragmentation distribution. Detailed fragmentation simulations at the
 lowest masses will be required to prove whether fragmentation can
 produce the VLM binary separation distribution observed.

\subsubsection{What effect does an ejection ``kick'' have on the binary distribution?}

Assuming (as has been suggested by \cite{rei02,bat02}) that eventually
a close multiple encounter ejects VLM ($<0.185 M_{\sun}$)
binaries, it would be interesting to see if such an ``ejected''
VLM binary distribution would have a maximum separation of $\sim16$ AU
as observed. The full dynamical modeling of the effect of a binary's
ejection from a mini-cluster is beyond the scope of this
paper. However, we can simply explore, as a ``toy-model'', the
possibility that the ejection process provides a hard differential
velocity kick of order $\sim 3$km/s to the two components of an ejected VLM
binary. We take the value of $\sim 3$ km/s since this is the
value calculated for the ejection velocity of on average more massive
single objects from a dissolving mini-cluster
\citep{ste98,rep02}. We assume, this may be of the right order of magnitude for
the differential velocity kick between the VLM members of the ejected
binary. We caution that this is only a simple toy-model, the true
differential velocity applied during the ejection can only be
calculated by detailed simulations of each ejection event.

In our
simple toy-model ejection only binaries with $V_{esc} \ga 3.0$ km/s
will survive the ejection process. Moreover, systems that are more tightly bound than  $V_{esc}=3.0$ will hardly be effected by such a kick in the tidal limit of \cite{wei87}. To survive in the general galactic
field we see (from Figure
\ref{fig3}) $V_{esc}\ga 0.6$ km/s. Hence one may expect today's
population of VLM binaries to all have $V_{esc_{VLM}} \ga (3.0+0.6)$
km/s.

It is very interesting to note that we ``observe'' (from figure
\ref{fig3}) $V_{esc}
\ga 3.8$ km/s for wide VLM binaries which is similar to the
``predicted'' value of $V_{esc_{VLM}} \ga 3.6$ km/s. Therefore, the
observed cut-off at 16 AU could be produced by a population of VLM
binaries where each system has been subjected to a differential kick of
$\sim3$ km/s in the ejection process. {\it We conclude that it is
possible to produce the observed distribution of semi-major
axes for VLM binaries by applying a strong differential velocity kick to each binary in
the distribution. However, it is still not clear if a VLM binary frequency of
$9-15\%$ could be produced by ejection, since ejection of a VLM tight binary may be a very rare event.}

\section {Future observations}

Future observations of all of these binaries should determine if there
is still Li present in their spectrum. The most useful Li observation
would spatially resolve both components in the visible
spectrum. Currently there is no visible wavelength AO systems capable
of guiding on a $V\sim 20$ source and obtaining resolutions better
than $0.1 \arcsec$ in the optical \citep{clo00,clo02d}. Therefore, we
will have to carry out such observations from space with HST/STIS
perhaps. Trigonometric parallax measurements should also be obtained
in the near future. Within the next few years one should also be able
to measure the masses of both components of several of these systems
as they complete a significant fraction of their orbit. Hence, further
observations of these systems are a high priority for the calibration
of the brown dwarf mass-age-luminosity relation. Continued searches
for new VLM binaries will help define this very interesting and
important binary population with greater precision.

\section {Conclusions about the VLM population in general}

 Based on all 34 VLM ($M_{tot}<0.185
M_{\sun}$) systems currently
known from this work and that of \cite{bas99,mar99,kor99,rei01a,lan01,pot02a,bur03,bou02}, we find the following general characteristics:

\begin{itemize}

\item VLM binaries are tight (peak separation $\sim4$ AU) with no
systems wider than 16 AU (this is $\sim 10$ times tighter than the slightly more massive M0-M4 binaries); 

%\item the widest VLM binaries seem to have $V_{esc} \sim 3.8$ km/s;

\item they
tend to have nearly equal mass companions ($q\sim0.9$) with no detected companions below $\sim 70\%$ of the primary's mass;

\item they have a
corrected binary frequency of $15\pm7 \%$ for spectral types M8-L0.5, and $9^{+15}_{-4}\%$ for T dwarf binaries \citep{bur03} for separations $a>2.6$ AU. This is less than the $32\pm9\%$ binary frequency of M0-M4 binaries \citep{fis92} and the $\sim50\%$  binarity of G dwarf binaries \citep{duq91} for similar separations.
\end{itemize}

We find the formation of such a VLM distribution to be problematic
with current simulations of binary star formation. 
\begin{itemize}

\item Ejecting VLM binaries from their formation
mini-clusters \citep{rep02} can likely produce the observed cut-off in
separation at 16 AU, but ejection simulations \citep{bat02} produce a binarity of $\la5\%$ which is smaller than the $\sim9-15\%$ binarity
observed.

\item If we ignore ejection and assume that these VLM
binaries are instead the residual (most massive) system remaining from
the dynamical decay of the mini-cluster one can likely also reproduce the
tight cut-off at 16 AU. But the strong bias to higher masses in the
final binary predicts that too few VLM binaries would be made in this fashion --assuming each component is picked randomly from the IMF
\citep{ste98, dur01}. 

\item We briefly look at a ``toy-model'' of a VLM binary population
produced by fragmentation. It appears that simply scaling an observed
fragmentation produced T-Tauri distribution to force a peak at 4 AU
produces systems with separations much greater than the 16 AU cut-off
observed. 
\end{itemize}

Hence, it is not obvious that the characteristics
of the VLM binary distribution are accurately predicted by any of
these methods. Future detailed modeling (taking into account circumstellar gas disks and their effects on dynamical interactions/ejections) will be required to see if
ejection or dynamical decay can be common enough to explain the
$\sim9-15\%$ binarity of VLM systems. Future detailed fragmentation
simulations of the smallest cores will be needed to discern if some type of
truncation occurs at low masses that limits VLM binaries to separations of less than
16 AU.

\acknowledgements

The Hokupa'a AO observations were supported by the University of
Hawaii AO group.  (D. Potter, O. Guyon, \& P. Baudoz). Support for
Hokupa'a comes from the National Science Foundation.  We thank the
anonymous referee for many detailed comments that led to an all round
better paper.  LMC acknowledges support by the AFOSR under
F49620-00-1-0294 and from NASA Origins NAG5-12086.  We would also like
to send a big {\it mahalo nui} to the Gemini operations staff.  These
results were based on observations obtained at the Gemini Observatory,
which is operated by the Association of Universities for Research in
Astronomy, Inc., under a cooperative agreement with the NSF on behalf
of the Gemini partnership: the National Science Foundation (United
States), the Particle Physics and Astronomy Research Council (United
Kingdom), the National Research Council (Canada), CONICYT (Chile), the
Australian Research Council (Australia), CNPq (Brazil) and CONICET
(Argentina).

%% Tables should be submitted one per page, so put a \clearpage before

%% deluxetable environment provided by the AASTeX package or the LaTeX
%% table environment.  Use of deluxetable is preferred.
%%

%% Three table samples follow, two marked up in the deluxetable environment,
%% one marked up as a LaTeX table.

%% In this first example, note that the \tabletypesize{}
%% command has been used to reduce the font size of the table.
%% Note also that the \label command needs to be placed 

\clearpage
\begin{deluxetable}{llllllllll}
\tabletypesize{\scriptsize}
\tablecaption{M8.0-L0.5 Stars Observed with No Likely Physical Companions
Between 0.1$\arcsec$-15$\arcsec$\label{tbl-0}}
\tablewidth{0pt}
\tablehead{
\colhead{2MASS} &
\colhead{other name} &
%\colhead{$J$} &
%\colhead{$H$} &
\colhead{K$_s$} &
\colhead{SpT} &
%\colhead{D$_{phot}$ (pc)} &
\colhead{Ref.} &
}
\startdata

2MASSW J0027559+221932 & LP 349-25 & \phn9.56 & M8.0        & 1\\
2MASSW J0140026+270150 & & 11.44 & M8.5        & 1\\
2MASSI\,\,\,\, J0149089+295613 & & 11.99 & M9.5        & 1\\
2MASSW J0253202+271333 & & 11.45 & M8.0        & 1\\
2MASSW J0320597+185423 & LP412-31 & 10.57 & M9.0        & 1\\
2MASSI\,\,\,\, J0335020+234235 & & 11.26 & M8.5        & 1\\
2MASSW J0350573+181806 & LP 413-53 & 11.76 & M9.0        & 1\\
2MASSW J0354013+231633 & & 11.97 & M8.5        & 1\\
2MASSW J0810586+142039 & & 11.61 & M9.0        & 1\\
2MASSW J0853361-032931 & LHS 2065 & \phn9.98  & M9.0        & 2\\
2MASSW J0928256+423054 & & 11.97 & M8.0        & 3\\
2MASSW J1019568+732408 & & 11.78 & M8.0 & 3\\
2MASSW J1124048+380805 & & 11.57 & M8.0 & 3\\
2MASSW J1224522-123835 & BRI 1222-1221 & 11.37 & M9.0        & 2\\
2MASSP\,\, J1309219-233035 & & 10.67 & M8.0 & 4\\
2MASSW J1403223+300755 & & 11.63 & M8.5        & 1\\
2MASSW J1421314+182740 & & 11.93 & M9.5        & 1\\
2MASSW J1444171+300214 & LP326-21 & 10.57 & M8.0        & 1\\
2MASSW J1457396+451716 & & 11.92 & M9.0        & 1\\
2MASSI\,\,\,\, J1501081+225001 & TVLM 513-46546& 10.72 & M8.5        &
5\\
2MASSW J1551066+645704 & & 11.73 & M8.5        & 1\\
2MASSW J1553199+140033 & & 11.85 & M9.0        & 1\\
2MASSW J1627279+810507 & & 11.87 & M9.0        & 1\\
2MASSW J1635192+422305 & & 11.80 & M8.0        & 1\\
2MASSW J1707183+643933 & & 11.39 & M9.0        & 1\\
2MASSW J1733189+463359 & & 11.86 & M9.5        & 1\\
2MASSI\,\,\,\, J2234138+235956 & & 11.81 & M9.5        & 1\\
2MASSI\,\,\,\, J2334394+193304 & & 11.64 & M8.0        & 1\\
2MASSW J2347368+270206 & & 12.00 & M9.0        & 1\\
2MASSW J2349489+122438 & LP 523-55 & 11.56 & M8.0        & 1\\
\enddata
\tablerefs{
(1)\cite{giz00}; (2)\cite{kir95}; (3)\cite{cru02}; (4)\cite{kir97}; (5)\cite{tin93}
}
\end{deluxetable}

\clearpage
\begin{deluxetable}{lllllll}
\tabletypesize{\scriptsize}
\tablecaption{The binary systems observed\label{tbl-1}}
\tablewidth{0pt}
\tablehead{
\colhead{System} &
\colhead{$\Delta J$} &
\colhead{$\Delta H$} &
\colhead{$\Delta K^{\prime}$} &
\colhead{Sep. ($\arcsec$)} &
\colhead{PA} &
\colhead{Age (Gyr)}
}
\startdata
%%2M1426\tablenotemark{b} & $0.78\pm 0.05$ & $0.70\pm 0.05$ &$0.65\pm 0.10$ &$0.57\pm 0.14$ & $0.152\pm0.006$ & $344.1\pm0.7^{\circ}$ & $0.8_{-0.2}^{+6.7}$\tablenotemark{b} & $23.6\pm 6.0$\\

2MASSW J0746425+200032\tablenotemark{a} & $0.60 \pm 0.20$ & $0.48 \pm0.15 $ &$0.44 \pm 0.15$ &$0.121 \pm0.008 $\tablenotemark{c} & $85.7 \pm1.45^{\circ} $\tablenotemark{c} & $5.0_{-4.15}^{+2.5}$\\ 
2MASSW J1047127+402644\tablenotemark{b} & $ 0.85\pm0.25 $ & $ 0.91\pm  0.20$ &$0.50 \pm 0.15$ &$ 0.122\pm0.008 $\tablenotemark{d} & $ 328.36\pm 3.75^{\circ} $\tablenotemark{d} & $5.0_{-4.4}^{+2.5}$\\
LHS 2397a\tablenotemark{g} & $3.83 \pm0.60 $ & $ 3.15\pm0.30 $ &$ 2.77\pm0.10 $ &$  0.207\pm0.007 $\tablenotemark{c} & $151.98 \pm1.20 ^{\circ} $\tablenotemark{c} & $7.2_{-5.2}^{+4.8}$\\
2MASSW J1127534+741107 & $0.33 \pm0.11 $ & $0.27 \pm 0.10$ &$ 0.25\pm 0.07$ &$0.246 \pm 0.008$\tablenotemark{c} & $80.23 \pm 1.72^{\circ} $\tablenotemark{c} & $3.0_{-2.4}^{+4.5}$\\
2MASSW J1311391+803222 & $ 0.13\pm 0.10$ & $ 0.15\pm 0.09$ &$ 0.14\pm 0.05$ &$ 0.267\pm0.006 $\tablenotemark{d} & $ 168.15\pm 0.48^{\circ} $\tablenotemark{d} & $5.0_{-4.4}^{+2.5}$\\
2MASSW J1426316+155701 & $ 0.78\pm0.05 $ & $ 0.70\pm0.05 $ &$ 0.65\pm0.10 $ &$ 0.152 \pm0.006 $\tablenotemark{e} & $344.1 \pm0.7 ^{\circ} $\tablenotemark{e} & $0.8_{-0.2}^{+6.7}$\\
2MASSW J2140293+162518 & $ 0.77\pm 0.05$ & $ 0.73\pm0.04 $ &$ 0.75\pm 0.04$ &$0.155 \pm  0.005$\tablenotemark{f} & $134.30 \pm 0.5^{\circ} $\tablenotemark{f} & $3.0_{-2.4}^{+4.5}$\\
2MASSW J2206228-204705 & $ 0.17\pm 0.04 $ & $ 0.08\pm 0.03 $ &$ 0.08\pm 0.03 $ &$0.168 \pm0.007 $\tablenotemark{f} & $68.2 \pm0.5 ^{\circ} $\tablenotemark{f} & $3.0_{-2.4}^{+4.5}$\\
2MASSW J2331016-040618 & $ 2.78\pm  0.04$ & $ 2.64\pm0.05 $ &$2.44 \pm0.03 $ &$ 0.573\pm0.008 $\tablenotemark{f} & $ 302.6\pm0.4 ^{\circ} $\tablenotemark{f} & $5.0_{-4.4}^{+2.5}$\\

%%2M2140 & $0.77\pm 0.05$ & $0.73\pm 0.04$ &$0.75\pm 0.04$ &$0.76\pm 0.13$ & $0.155\pm0.005$ & $134.3\pm0.5^{\circ}$ & $3.0_{-2.4}^{+4.5}$ & $23.9\pm 6.0$\\
%%2M2206 &$0.17\pm 0.04$ & $0.08\pm 0.04$ &$0.08\pm 0.03$ &$0.08\pm 0.14$ & $0.168\pm0.007$ & $68.2\pm0.5^{\circ}$ & $3.0_{-2.4}^{+4.5}$ & $24.68\pm 6.8$\\
%%2M2331 &$2.78\pm 0.04$ & $2.64\pm 0.05$ &$2.44\pm 0.03$ &$2.38\pm 0.16$ & $0.573\pm0.008$ & $302.6\pm0.4^{\circ}$ & $5.0_{-4.4}^{+2.5}$ & $25.2\pm 6.8$\\
\enddata
\tablenotetext{a}{discovery paper \cite{rei01a}}
\tablenotetext{b}{discovery paper \cite{rei02}}
\tablenotetext{c}{observations made on Feb 7, 2002 UT}
\tablenotetext{d}{observations made on April 25, 2002 UT}
\tablenotetext{e}{observations made on June 20, 2001 UT}
\tablenotetext{f}{observations made on Sept 20, 2001 UT}
\tablenotetext{g}{discovery paper \cite{fre02}}
%%2M1426 observations made on June 20, 2001 \cite{clo02b}; the young age of 2M1426 is motivated in \cite{clo02b}}
\end{deluxetable}

\clearpage
\begin{deluxetable}{llllllllll}
\tabletypesize{\scriptsize}	
\tablecaption{Summary of the new binaries' A \& B components \label{tbl-2}}
\tablewidth{0pt}
\tablehead{
\colhead{Name} &
\colhead{$J$} &
\colhead{$H$} &
\colhead{$Ks$} &
\colhead{$M_{K_{s}}$\tablenotemark{c}} &
\colhead{SpT\tablenotemark{a}}&
\colhead{Est. Mass\tablenotemark{b}} &
\colhead{Est. D (pc)\tablenotemark{c}}&
\colhead{Sep. (AU)} &
\colhead{P (yr)\tablenotemark{d}}
}
\startdata
2M0746A &$12.23 \pm 0.083 $ &$11.53 \pm0.07 $ &$11.04 \pm0.07 $ &$10.58 \pm0.07 $ & L0.5 &$0.082_{-0.004}^{+0.001}$ & $12.34\pm0.05\tablenotemark{e} $&$ 1.47 \pm0.11 $ &$11_{-2}^{+3}$\\
2M0746B &$12.83 \pm0.21  $ &$12.01 \pm0.16 $ &$ 11.48\pm0.16 $ &$11.0 \pm0.17 $ & L2 &$0.078_{-0.009}^{+0.001}$ & & & \\

2M1047A &$ 12.85\pm0.08  $ &$ 12.09\pm0.07 $ &$ 11.80\pm0.07 $ &$ 9.95\pm0.39 $  &M8  &$0.092_{-0.012 }^{+0.009}$ & $ 23.4\pm4.3  $&$2.9  \pm0.6 $ &$ 11_{-3 }^{+5 }$\\
2M1047B &$13.70 \pm0.26  $ &$ 13.00\pm0.21 $ &$ 12.31\pm0.16 $ &$ 10.46\pm0.42 $ &L0 &$0.084_{-0.016 }^{+0.006 }$ & & & \\

LHS 2397aA &$ 11.86\pm0.05 $ &$ 11.32\pm0.05 $ &$ 10.80\pm 0.03 $ &$10.03 \pm0.09 $ &M8  &$0.090_{-0.001 }^{+0.004}$ & $ 14.3\pm0.4\tablenotemark{e}  $&$ 3.86 \pm0.18 $ &$ 22_{-3 }^{+3 }$\\
LHS 2397aB &$15.69 \pm 0.60 $ &$ 14.47\pm 0.30$ &$13.57 \pm0.10 $ &$ 12.80\pm 0.12$ &L7.5 &$ 0.068_{-0.007 }^{+0.001 }$ & & & \\

2M1127A &$ 13.66\pm0.06 $ &$ 12.99\pm 0.05$ &$12.60 \pm0.05 $ &$ 9.96\pm0.37 $ &M8 &$ 0.092_{-0.012 }^{+0.009 }$ & $ 33.67\pm 5.9 $&$  8.31\pm1.49 $ &$ 57_{-15 }^{+26 }$\\
2M1127B &$ 13.99\pm0.13  $ &$ 13.26\pm0.11 $ &$ 12.85\pm  0.08$ &$ 10.21\pm 0.38$ &M9 &$ 0.087_{-0.013 }^{+0.007 }$ & & & \\

2M1311A &$ 13.49\pm 0.06$ &$ 12.82\pm0.05 $ &$ 12.39\pm0.04 $ &$ 10.07\pm0.37 $ & M8.5 &$0.089_{-0.011 }^{+0.008 }$ & $29.1 \pm 5.1 $&$  7.7\pm 1.3 $ &$ 51_{-13 }^{+24 }$\\
2M1311B &$ 13.62\pm0.11  $ &$ 12.97\pm0.10 $ &$ 12.53\pm0.07 $ &$ 10.21\pm 0.38$ &M9 &$ 0.087_{-0.013 }^{+0.007 }$ & & & \\

2M1426A &$  13.30\pm0.04 $ &$  12.63\pm0.04 $ &$ 12.18\pm0.05 $ &$ 10.09\pm0.37 $ & M8.5 &$0.088_{-0.010 }^{+0.009 }$ & $ 26.1\pm4.5  $&$  3.9\pm0.7 $ &$ 19_{-5 }^{+9 }$\\
2M1426B &$ 14.08\pm 0.06 $ &$ 13.33\pm0.06 $ &$ 12.83\pm 0.11$ &$ 10.74\pm0.38 $ &L1 &$ 0.076_{-0.014 }^{+0.008 }$ & & & \\

2M2140A &$ 13.37\pm0.04 $ &$ 12.71\pm0.04 $ &$ 12.22\pm0.04 $ &$ 10.17\pm0.37 $ &M9  &$0.088_{-0.012 }^{+0.007 }$ & $ 25.6\pm4.4  $&$  3.9\pm0.6 $ &$ 19_{-5}^{+9 }$\\
2M2140B &$ 14.14\pm0.06  $ &$ 13.44\pm0.05 $ &$ 12.97\pm0.05 $ &$ 10.92\pm 0.37 $ &L2 &$ 0.078_{-0.020 }^{+0.004 }$ & & & \\

2M2206A &$  13.10\pm0.04 $ &$ 12.46\pm 0.04$ &$ 12.06\pm0.04 $ &$ 9.93\pm 0.37$ & M8 &$0.092_{-0.011 }^{+0.009 }$ & $ 26.7\pm 4.5 $&$  4.4\pm0.7 $ &$ 22_{-5 }^{+10 }$\\
2M2206B &$  13.27\pm 0.05 $ &$ 12.54\pm 0.05$ &$ 12.14\pm0.05 $ &$ 10.01\pm 0.37$ &M8 &$ 0.091_{-0.011 }^{+0.008 }$ & & & \\

2M2331A &$ 13.02\pm 0.04$ &$ 12.38\pm 0.04 $ &$ 12.03\pm 0.04$ &$ 9.93\pm0.08 $ & M8.0 &$0.093_{-0.002 }^{+0.002 }$ & $ 26.2\pm 0.6\tablenotemark{e} $&$  15.0\pm0.37$ &$ 159_{-39 }^{+70 }$\\
2M2331B &$ 15.80\pm0.05  $ &$ 15.02\pm 0.06$ &$ 14.47\pm0.05 $& $12.37 \pm 0.12$  & L7 &$ 0.067_{-0.013 }^{+0.002 }$ & & & \\

%%2M1426A &$13.36\pm 0.06$ &$12.63\pm 0.05$ &$12.20\pm 0.07$ &$12.07\pm 0.08$ & $1.16\pm0.12$& M8.5 & $0.083_{-0.014}^{+0.010}$& $3.6\pm 0.9$& $17_{-7}^{+10}$\\

%%2M1426B &$14.13\pm 0.06$ &$13.34\pm 0.10$ &$12.80\pm 0.14$ &$12.64\pm 0.14$ & $1.33\pm0.12$& L1 & $0.075_{-0.020}^{+0.009}$ & & \\

%%2M2140A &$13.37\pm 0.06$ &$12.72\pm 0.04$ &$12.22\pm 0.04$ &$12.07\pm 0.09$ & $1.15\pm0.12$& M8.5 & $0.087_{-0.017}^{+0.008}$& $3.7\pm 0.9$& $18_{-7}^{+10}$\\
%%2M2140B & $14.15\pm 0.06$ & $13.44\pm 0.04$ &$12.97\pm 0.04$ &$12.83\pm 0.09$ & $1.19 \pm 0.14$& L0 & $0.075_{-0.018}^{+0.007}$ & &  \\

%%2M2206A &$13.12\pm 0.06$ &$12.46\pm 0.06$ &$12.06\pm 0.05$ &$11.94\pm 0.08$ & $1.06\pm0.12$& M8.0 & $0.090_{-0.014}^{+0.008}$& $4.1\pm 1.1$& $20_{-7}^{+9}$\\
%%2M2206B & $13.27\pm 0.06$ & $12.54\pm 0.06$ &$12.14\pm 0.05$ &$12.02\pm 0.08$ & $1.13 \pm 0.14$& M8.5 & $0.088_{-0.014}^{+0.008}$ & &  \\

%%2M2331A &$13.08\pm 0.04$ &$12.38\pm 0.04$ &$12.04\pm 0.04$ &$11.94\pm 0.07$ & $1.04\pm0.12$& M8.0 & $0.091_{-0.013}^{+0.008}$& $14.4\pm 3.9$& $139_{-57}^{+86}$\\
%%2M2331B & $15.86\pm 0.06$ & $15.03\pm 0.06$ &$14.48\pm 0.06$ &$14.32\pm 0.10$ & $1.38 \pm 0.14$& L3 & $0.062_{-0.020}^{+0.010}$ & &  \\

\enddata
\tablenotetext{a}{Spectral types estimated by $3.97$x$M_{Ks}-31.67$ \citep{dah02,sie02} with $\pm1.5$ spectral subclasses of error in these estimates (note a value of SpT=10 is defined as L0 in the above equation).}
\tablenotetext{b}{Masses (in solar units) from the models of \cite{cha00} --see Figure \ref{fig2}.}
\tablenotetext{c}{Photometric distances estimated for the primaries by $M_{Ks} = 7.71+2.14(J-Ks)$ which is valid for $M7<SpT<L1$ \citep{sie02}.}
\tablenotetext{d}{Periods estimated assuming face-on circular orbits. In the cases of 2M0746 and LHS 2397a, the larger of the 2 observed separations (2.7 AU \& 3.86 AU from \cite{rei01a} and \cite{fre02}, respectively) have been used to more accurately estimate the period.} 
\tablenotetext{e}{These systems have trigonometric parallaxes.}
\end{deluxetable}

%%\clearpage
%%\begin{deluxetable}{lll}
%%\tabletypesize{\scriptsize}
%%\tablecaption{Summary of the 2M1426 A \& B components \label{tbl-2}}
%%\tablewidth{0pt}
%%\tablehead{
%%\colhead{Parameter} &
%%\colhead{2M1426A} &
%%\colhead{2M1426B}
%%}
%%\startdata
%%H mag   & $12.63\pm 0.05$ & $13.34\pm 0.10$ \\
%%Ks mag  & $12.20\pm 0.07$ & $12.80\pm 0.14$ \\
%%K mag   & $12.16\pm 0.05$ & $12.75\pm 0.10$ \\
%%H-K color & $0.47\pm 0.08$ & $0.59\pm 0.14$  \\
%%Est. SpT & M8.5            & L1-L3              \\
%%Mass & $0.074^{+0.005}_{-0.011} M_\odot$ & $0.066^{+0.006}_{-0.015} M_\odot$ \\
%%\enddata
%%end{deluxetable}

\clearpage
\begin{deluxetable}{lllllll}
\tabletypesize{\scriptsize}
\tablecaption{All Known Resolved VLM Binaries\tablenotemark{e} \label{tbl-3}}
\tablewidth{0pt}
\tablehead{
\colhead{Name} &
\colhead{Sep.} &
\colhead{Est.} &
\colhead{Est. $M_A$}&
\colhead{Est. $M_B$}&
\colhead{Est. Period\tablenotemark{b}} &
\colhead{Ref.\tablenotemark{c}} \\
\colhead{} &
\colhead{AU} &
\colhead{$SpT_A/SpT_B$} &
\colhead{$M_{\sun}$} &
\colhead{$M_{\sun}$} &
\colhead{yr} &
\colhead{} 
}  
\startdata
PPL 15\tablenotemark{a}   &  0.03 & M7/M8 & 0.07 & 0.06 & 5.8 days & 0 \\
Gl 569B  &  1.0  & M8.5/M9.0 & 0.063 & 0.06 & 3 & 1,2 \\
SDSS 2335583-001304 & 1.1? & L1?/L4? & 0.079 & 0.074 & 3 & 12 \\ 
2MASSW J1112256+354813 & 1.5 & L4/L6 & 0.073 & 0.070 & 5 & 12 \\
2MASSI J1534498-295227 & 1.8 & T5.5/T5.5 & 0.05 & 0.05 & 8 & 9\\  
2MASSW J0856479+223518 & 2.0 & L5?/L8? & 0.071 & 0.064 & 8 & 12 \\
DENIS-P J185950.9-370632 & 2.0 & L0/L3 & 0.084 & 0.076 & 7 & 12 \\  
HD130948B &  2.4  & L2/L2 & 0.07 & 0.06 & 10 & 3  \\
2MASSW J0746425+200032   &  2.7  & L0.5/L2 & 0.082 & 0.078 & 12 & 4 \& this paper  \\
2MASSW J1047127+402644   &  2.7  & M8/L0 & 0.092 & 0.084 & 11 & 8 \& this paper\\
DENIS-PJ035726.9-441730 & 2.8    & L2/L4 & 0.078 & 0.074 & 12 & 12\\
2MASSW J0920122+351742   &  3.2  & L6.5/L7 & 0.068 & 0.068 & 16  & 4 \\
LP415-20 & 3.5 & M7/M9.5 & 0.095 & 0.079 & 15 & 11, this survey \\
2MASSW J1728114+394859 & 3.7 & L7/L8 & 0.069 & 0.066 & 19 & 12 \\ 
LHS 2397a & 3.9 & M8/L7.5 & 0.090 & 0.068 & 22 & 7,this paper\\
2MASSW J1426316+155701   &  3.9  & M8.5/L1 & 0.088 & 0.076 & 19 & this paper\\
2MASSW J2140293+162518   &  3.9 & M9/L2 & 0.092 & 0.078 & 22 & this paper \\
2MASSW J2206228-204705   &  4.4  & M8/M8 & 0.092 & 0.092 & 22 & this paper\\ 
2MASSs J0850359+105716   &  4.4  & L6/L8 & 0.05 & 0.04 & 30 & 4 \\
2MASSW J1750129+442404   &  4.8  & M7.5/L0 & 0.095 & 0.084 & 25 & 11, this survey \\
DENIS-P J1228.2-1547 & 4.9 & L5/L5 & 0.05 & 0.05 & 34 &  10 \\
2MASSW J1600054+170832 & 5.0 & L1/L3 & 0.078 & 0.075 & 29 & 12\\
2MASSW J1239272+551537 & 5.1 & L5/L5 & 0.071 & 0.071 & 31 & 12 \\
%2MASSI J1225543-273946\tablenotemark{f} ? & 5.2 &T6/T8 ? & 0.045 & 0.03 & 43 & 9 \\
2MASSI J1553022+153236 & 5.2 & T7/T7.5 & 0.040 & 0.035 & 43 & 9 \\ 
2MASSW J1146345+223053	& 7.6 & L3/L4 & 0.055 & 0.055 & 63 & 6 \\
2MASSW J1311391+803222  & 7.7 & M8.5/M9 & 0.089 & 0.087 & 51 & this paper\\
2MASSW J1127534+741107  & 8.3 & M8/M9 & 0.092 & 0.087 & 57 & this paper\\
LP475-855 & 8.3 & M7.5/M9.5 & 0.091 & 0.080 & 58 & 11, this survey \\ 
DENIS-P J0205.4-1159 & 9.2 & L7/L7 & 0.07 & 0.07 & 75 & 6 \\
2MASSW J2101349+175611 & 9.6 & L7/L8 & 0.068 & 0.065 & 82 & 12\\
2MASSW J2147436+143131 & 10.4 & L0/L2 & 0.084 & 0.078 & 83 & 12 \\
2MASSW J1449378+235537 & 11.7 & L0/L3 & 0.084 & 0.075 & 100 & 12\\
DENIS-P J144137.3-094559 & 13.5 & L1/L1 & 0.079 & 0.079 & 124 & 12\\
2MASSW J2331016-040618   & 15.0 & M8.0/L7 & 0.093 & 0.067 & 159 & this paper \\
\enddata
\tablenotetext{a}{PPL 15 is a spectroscopic binary \citep{bas99}}
\tablenotetext{b}{This ``period'' is simply an estimate assuming a face-on circular orbit}
\tablenotetext{c}{REFERENCES--(0)\cite{bas99}; (1) \cite{ken01}; (2) \cite{lan01}; (3) \cite{pot02a}; (4)\cite{rei01a}; (5) \cite{mar99}; (6) \cite{kor99}; (7) \cite{fre02}; (8) \cite{rei02}; (9) \cite{bur03}; (10) \cite{mar99}; (11) \cite{sie02}; (12) \cite{bou02}}
\tablenotetext{d}{Gl 569B and HD130948B are binary brown dwarfs that orbit normal stars, for AO observations these bright primary stars were guided on, not the brown dwarfs}
\tablenotetext{e}{We define VLM binaries as 2 star systems where $M_{tot}<0.185 M_{\sun}$. Very young evolving systems (like GG TauBaBb \citep{whi01}) are not included, nor are over-luminous systems which are not resolved into binaries.}
%\tablenotetext{f}{2MASSI J1225543-273946B may be just a cosmic ray \citep{bur03}}
\end{deluxetable}

\clearpage

\clearpage

%%\figcaption[f2.eps]{
%%The 2MASSJ 1426316+155701 binary at H band resolution
%%0.131\arcsec$. In Figure b we show the sytem after being LUCY
%%restored to $0.080\arcsec$ resolution. Only astrometry was derived
%%from the deconvolved images.\label{fig1}}

\clearpage

\clearpage

\clearpage

\clearpage

\clearpage

\clearpage

\clearpage

\clearpage

\clearpage

\clearpage

\clearpage

\clearpage

\clearpage

\clearpage

\clearpage

\clearpage

\end{document}